\documentclass[aps,rmp,twocolumn,twoside]{revtex4}
\usepackage{latexsym}
\usepackage{graphicx}
\usepackage{fancyhdr}
\usepackage{amsbsy}
\usepackage{color}
\def\lsim{\mathrel{\raise3pt\hbox to 8pt{\raise -6pt\hbox{$\sim$}\hss {$<$}}}}
\newcommand{\be}{\begin{equation}}
\newcommand{\ee}{\end{equation}}
\newcommand{\bea}{\begin{eqnarray}}
\newcommand{\eea}{\end{eqnarray}}

\newcommand{\lc}{^-\!\!\!\!\lambda_C}
\def\bop#1{\setbox0=\hbox{$#1M$}\mkern1.5mu
\vbox{\hrule height0pt depth.04\ht0
\hbox{\vrule width.04\ht0 height.9\ht0 \kern.9\ht0
\vrule width.04\ht0}\hrule height.04\ht0}\mkern1.5mu}

\begin{document}


\title{Photon and Graviton Mass Limits }
\author{Alfred Scharff Goldhaber$^{*,\dagger}$ and Michael Martin
Nieto$^{\dagger}$
\\}

\affiliation{
$^{*}$C. N. Yang Institute for Theoretical Physics, SUNY Stony Brook, NY
11794-3840 USA
\smallskip
\\
and
\smallskip
\\
$^{\dagger}$Theoretical Division (MS B285), Los Alamos National Laboratory, Los
Alamos, NM 87545 USA\\}

\begin{abstract}

Efforts to place limits on deviations from canonical formulations of electromagnetism and gravity have 
probed length scales increasing dramatically over time.
Historically, these studies have passed through three stages:  (1) Testing the power in the inverse-square laws of Newton and
Coulomb,  (2) Seeking a nonzero value for the rest mass of photon or graviton, (3) Considering more degrees of freedom, allowing mass while preserving explicit
gauge or general-coordinate invariance.  Since our previous review the lower limit on the photon Compton wavelength has improved by four orders of magnitude, to about one astronomical unit, and rapid current progress in astronomy makes further advance likely.  For gravity there have been vigorous debates about even the concept of graviton rest mass.  Meanwhile there are striking observations of astronomical motions that do not fit Einstein gravity with visible sources.  ``Cold dark matter" (slow, invisible classical particles) fits well at large scales. ``Modified Newtonian dynamics" provides the best phenomenology at galactic scales. Satisfying this phenomenology is a requirement if dark matter, perhaps as invisible classical fields, could be correct here too.  ``Dark energy" {\it might} be explained by a graviton-mass-like effect, with associated Compton wavelength comparable to the radius of the visible universe.  We summarize  significant mass limits in a table.

\end{abstract}

\today

\maketitle


\tableofcontents



\section{Introduction}

Photons and gravitons are the only known free particles whose rest masses may be exactly zero.\footnote{
Gluons, the gauge particles of quantum chromodynamics, are believed to have no bare mass.  However, they are not seen in isolation, meaning they cannot be observed as free particles.
}
This bald statement covers a rich and complex history, from Newton and Gauss, through Maxwell and Einstein, and even up to the present.   During that development, the matrix of interlocking concepts surrounding the notions of photon and graviton rest mass, or more generally, long-distance, low-frequency deviations from Maxwell electrodynamics and Einstein gravity, has become increasingly elaborate.

There are many similarities between the photon and graviton  cases, but also striking differences.  We literally see photons all the time, as only a few photons of visible light are enough to activate one `pixel' in a human retina.  Besides that, conspicuous electromagnetic wave phenomena play an enormous role in modern physics.

For gravity the situation is radically different.  Even gravitational waves are in a situation analogous to that of neutrinos during the first 25 years after their proposal by Pauli, when their emission could be inferred from loss of energy, momentum, and angular momentum in beta decay, but they hadn't yet been detected in an absorption experiment.  Binary pulsar systems exhibit energy loss well accounted-for by radiation of gravitational waves, but experiments underway to detect absorption of such waves have not yet achieved positive results.   Even if this were accomplished sometime soon, the chances  of ever detecting
individual quanta -- gravitons -- seem remote indeed, because graviton coupling to matter is so enormously weak.

These are not the only differences.  For electrodynamics, the history has been one of increasingly sensitive null experiments giving increasingly stringent constraints on a 
possible mass.   Nevertheless, theories describing such a mass seem well-developed and consistent, even if not esthetically appealing.  On the other hand, for gravity, there {\it are} long-distance effects which some argue provide evidence supporting modification of Einstein's formulation.  At the same time, the theoretical basis for a gravitational phenomenon analogous to photon mass has come under severe attack. All this means that, with respect to the issue of deviations, currently there is much more dynamism for gravity than for electrodynamics.

Even though quantum physics gave shape to the concept of mass for
electrodynamics and gravitation, the obvious implication of a dispersion in velocity  with energy for field quanta, or even for waves, is beyond our capacity to detect with methods identified so far.  This is thanks to the very strong limits already obtained based on essentially static fields.  Thus, the domain of potentially interesting  experiment and observation for mass or ``mass-like" effects indeed is restricted to the long-distance and low-frequency scales already mentioned.


\subsection{How to test a theory}

Let us begin by seeking a broad perspective on what it means to probe, not
merely the validity but also the accuracy of a theory.   The canonical view of theory-testing  is that one tries to falsify the theory:   One compares its predictions with experiment and observation. The predictions use input data, for example initial values of certain parameters, which then are translated by the theory into predictions of new data.  If these predicted  data  agree with observation within experimental uncertainties (and sometimes also uncertainties in application of the theory), then the theory has, for the moment, passed the test. One may continue to look for failures in new domains of application, even if  the incentive for doing so  declines with time.

Of course, without strong `ground rules' it is impossible to falsify a theory, because one almost always can find explanations for a failure.  So, in fact no scientific theory either may be disproved or proved in a completely rigorous way; everything always is provisional, and continual skepticism always is in order.  However, based on a strong pattern of success a  theory can earn trust at least as great as in any other aspect of human inquiry.

The above is an essential, but we believe only partial, view of how theories gain conviction.  At least three important additional factors may help to achieve that result.

First, a striking, even implausible, prediction is borne out by experiment or observation.
Examples of this include `Poisson's spot', Poisson's devastating attack on the notion that light is a wave phenomenon, because this would require that the shadow of a circular obstacle have a bright spot in its center.   The discovery of the spot by Arago provided the conceptual equivalent of a judo maneuver, using the opponent's own impulse to overcome him.  Another example is the assertion by Appelquist and Politzer in the summer of 1974 that the existence of a heavy quark carrying the quantum number `charm' would imply the existence of a positronium-like spectrum, meaning very sharp resonances in electron-positron scattering.  When the J/$\psi$ was discovered at BNL and SLAC in November of that year, the outlandish prediction of Appelquist and Politzer suddenly was the best explanation, in good part because it was the only one that had been stated boldly beforehand (though only published afterwards).

A second way in which a theory gains credence is by fecundity:   People see ways to apply the idea in other contexts.  If many such applications are fruitful, then by the time initial experimental verification is rechecked there may be little interest, because the theory already has become a foundation stone for a whole array of applications.
An example from our subject here is the transfer of the $1/R^2$ force law from gravity to electricity, in a speculative leap during the 1700s.  Of course, it was not this transfer which gave Newton's gravity its great authority, but rather the enormous number of precise and successful predictions of his theory -- fecundity in the original, literally astronomical domain.

Closely related to the above is a third feature, connectivity.   If many closely neighboring subjects are described by connecting theoretical concepts, then the theoretical structure acquires a robustness which makes it increasingly hard -- though certainly never impossible -- to overturn.

The latter two concepts fit very well with Thomas Kuhn's epilogue to his magnum opus {\it The Structure of Scientific Revolutions} \cite{kuhn}, in which he muses that the best description of scientific development may be as an evolutionary process.   For biological evolution, both  fecundity and reinforcing connections play decisive roles in how things happen.  It seems to us, as it did to Kuhn, that the same is often true for {\it ideas} in all science, including physics.

To the extent that observational errors can be ruled out, whenever discrepancies appear between theory and experiment one is compelled to contemplate the possibility that new, or at  least  previously unaccounted-for, physics is contributing to the phenomena.  A classic case is the famous solar-neutrino puzzle.  At first, presumed errors in the actual measurements themselves were widely and caustically viewed as the problem.  Later critiques focused more on  consideration of 
possible 
errors in models of processes in the Sun and, perhaps more creatively, on potential new physics modifying the simplest picture  of neutrino propagation from source to detector.   By now there is overwhelming evidence that neutrino mixing, a modification of neutrino propagation, accounts precisely for the observed rate of neutrino observation on Earth, confirming the basic validity of  early work on both solar modeling and
neutrino detection.

A more refined question about confirming a theory is:   How does one quantify limits on deviations from the theory?   Now it becomes necessary to specify some form of  deviation that depends on certain parameters.  Then experimental uncertainties can be translated into quantitative limits on these parameters.   As a theory evolves, the favored choice for  an interesting form of deviation may change. Of course, at any point new observations contradicting even well-established theoretical predictions can reopen issues that might have seemed settled


\subsection{Photons}

Our earlier review \cite{GN}  described the state of theory for accommodating a non-zero photon mass, and the state of observation and experiment giving limits on the mass at that time.   Since then there have been significant theoretical developments (Section \ref{EM}), as well as advances in experimental approach and precision (Section \ref{EMtests}).

The issue of possible long-range  deviations from the existing theory of
electrodynamics  long remained purely a matter of choosing, and then setting limits on, parameters of a possible deviation.
Originally, in analogy to studies of Newton's Law for gravity, deviations  in power of radius from that in the inverse square law were used.  20th-century relativistic wave equations led to discussion in terms of a finite rest mass of the photon, thus introducing a length scale.
Today one can consider a more sophisticated approach 
explicitly incorporating gauge 
invariance, even when describing nonzero photon mass, 
by including more couplings. 

We shall not reprise in detail the theoretical paradigms and experimental
details discussed in Ref. \cite{GN} for the photon.  Rather, we refer the reader to that work for an introduction, and focus here on elucidating more recent advances.   Also, since the publication of \cite{GN}, there have been other works which have summarized specific aspects of the photon mass  \cite{early}-\cite{pdg}.  These can be consulted especially for experimental summaries.  In particular, Byrne \cite{early} concentrated on astrophysical limits, Tu and Luo \cite{metro}  on laboratory limits based on tests of Coulomb's Law, and, joined by Gillies, on all experimental tests  \cite{luorev}.  Okun, in a compact review \cite{okun}, gives some  interesting early history on the concept of the ``photon" in quantum
mechanics. He also gives details on what the Russian school accomplished during the earlier period. 
Vainshtein has presented a concise, and  insightful, perspective on theoretical issues relating to mass both for photon and graviton \cite{Vainshtein:2006ix}.  

Of course the types of deviation we discuss in this review do not cover all possibilities.   In particular, the Maxwell equations are linear in  the electromagnetic field strengths.  This does not mean that all phenomena are linear, because the coupling between fields and currents allows back-reaction and thus nonlinearity.  Nevertheless, the linearity of the equations is an especially simple feature.   Higher-order terms in the field strengths clearly are an interesting possibility, but they are intrinsically tied to short-distance modifications of the theory, rather than the long-distance deviations emphasized here.  The reason is that the nonlinear terms become more important as the field strengths increase, meaning that the numbers of flux lines per unit area increase, clearly
a phenomenon associated with short distance scales.

For completeness, we  briefly mention here discussions in the literature that go in the direction of nonlinearity.   Born and Infeld \cite{Infeld} introduced the notion of nonlinear damping of electromagnetic fields, precisely to cope with the short-distance singularities of the classical linear theory.   Their approach was pursued by a number of investigators over many years, as reviewed by Plebanski \cite{Plebanski}.  More recently their ideas have been revived because the kind of structure they discussed arises naturally in string  theory,\footnote{
We refer a number of times in this paper to string theory.   Even though  incomplete at this stage, one way it can be viewed is as a natural progression from field theories such as electrodynamics and general relativity.  In a number of cases  mathematical patterns found in string theory have led to discoveries of related patterns in field theory.}
as reviewed by Tseytlin and by Gibbons \cite{Tseytlin:1999dj,Gibbons:2001gy}.  A second approach to nonlinearity was introduced by Heisenberg and Euler \cite{H-E}, who observed that what we now would call virtual creation of electron-positron pairs leads inevitably to an extra term in the Maxwell equations that is cubic
in field strengths.  This work also has a living legacy, as reviewed recently by Dunne \cite{Dunne:2004nc}.


\subsection{Gravitons}

The scientific question in gravity most naturally related to photon mass is the issue of a possible graviton mass.  
(With the exception of Vainshtein's article \cite{Vainshtein:2006ix} mentioned above, devoted mainly to theoretical aspects, 
there apparently has never been a widely-circulated review on this topic.)
Its study progressed more slowly than the photon-mass issue, at least in part because gravity is so weak that even today classical gravitational waves have not been detected directly.  Further, gravitons, regardless of their mass, seem beyond the possibility of detection in the foreseeable future.

We proceed to discuss theoretical issues (Section \ref{grav}) and observations (Section \ref{gravmass}) for the case of classical gravity.   There are several important contrasts between the photon and graviton cases.  First, from a  theoretical point of view the possibility of nonzero graviton mass is open to question.  This makes what is  a relatively straightforward discussion for photons much more problematic for gravitons.  Secondly, there is a highly developed formalism for seeking to measure deviations of gravity from Einstein's General Theory of Relativity -- the parametrized post-Newtonian [PPN] expansion.

In this framework there have been many measurements, principally under weak-field conditions, both for low-velocity and high-velocity phenomena, to test for deviations.  As with photon mass, none of these measurements to date have produced ``unexpected-physics" results, only increasingly stringent limits on departures from Einstein gravity.

However, there is another important distinction.  Two sets of characteristic phenomena show significant departures from Einstein gravity with the matter sources being only familiar ``visible" matter -- stars, hot gases, and photons.  The first, indicated already by observations in the 1930s, and much more definitely in the 1970s, has been labeled  ``dark matter."  Trajectories of visible objects (including the most visible of all -- light itself) seem to be bent more than would be expected if the only sources for gravity were pieces of visible matter.   In principle, a possible explanation for this could be long-range modifications of Newtonian and Einsteinian gravity,  but of a type very different from what would be called graviton mass.  The second departure, discovered much more recently, is an accelerated expansion of the universe, possibly described by the presence of another sort of invisible source, ``dark energy."  We discuss these issues also in Section \ref{grav}.


\subsection{Perspectives on gauge and general-coordinate invariance}

This review, then, can be considered an evolution of our earlier review on the mass of the photon \cite{GN}.  Here we discuss current understanding and ideas on the masses of the photon and graviton, in light of  developments in theoretical and experimental physics over the past decades.

Because in principle there is no end to the  types of deviation that could be contemplated, we need some restrictions.   Our chosen class of deviations will be ones  that explicitly obey abstract ``symmetries:" 
gauge invariance in the case of electrodynamics, and general-coordinate invariance in the case of gravity.\footnote{In 
the case of string theory, the corresponding property is called ``reparametrization invariance."
}   

The distinction between these invariances and familiar global symmetries such as rotation or translation invariance is that not only the action but in fact {\it all} observables are invariant.   Thus the invariance is with respect to description rather than physical transformation.
At first sight it might appear that making such an invariance explicit would be useless, because no matter what the dynamics this always should be possible.   However, history shows that the invariances can be most fruitful:   First of all, electrodynamics and Einstein gravity are examples of {\it minimal} theories exhibiting such invariance, and hence are essentially unique.    Secondly, even for non-minimal versions with extra couplings, keeping the invariances explicit, while not constraining the physical content of the theory, can be most helpful in carrying through computations. 
They then play a  similar role to the ``check bit" at each stage in a numerical computation, which also adds nothing to the content, but can guide calculations and flag errors.  

It might seem that such invariance excludes a mass for the Proca photon or the graviton.  However, as we shall 
discuss, the Higgs mechanism can `hide' 
an invariance, 
which nevertheless remains unbroken.   For the photon that becomes equivalent in a certain limit to the fixed Proca mass, which therefore needn't break gauge invariance after all.


\section{Electromagnetic theories}
\label{EM}

As indicated in the introduction, one can mark three stages in the search for long-range deviations from  electrodynamics, of which the first two were described in our previous review \cite{GN}.
That article  appeared just as a `sea change' in the theoretical picture of physics was beginning to emerge, the notion that gauge theories and gauge invariance might underlie not only electromagnetism and gravity but also weak and strong interactions.  Nevertheless, until quite recently there had been surprisingly little discussion of the new perspective in the context of photon mass.


\subsection{Power-law deviation from Coulomb's form}

The first stage, as recounted in \cite{GN}, focused on the inverse-square force for the interactions of electric charges or magnetic poles.
The guess was that the strength of the electric force along the line between two charges would be similar to Newton's Law,
\begin{equation}
F=\frac{kq_1q_2}{r^2}   .
\end{equation}

Early experimenters chose to parametrize possible deviations from this form by what today we would call 
preserving the scaling or self-similariy exhibited by Newton's law. 
Presumably  this was because they had no framework to choose a particular length parameter instead.  Therefore they looked for modifications of the form
\begin{equation}
F= \frac{kq_1q_2}{r^{2+\alpha}}    ,
\end{equation}
and sought limits for a possible shift in power  $\alpha$ from the
inverse-square.  This early history, which started before Coulomb (although Coulomb eventually received credit for the law) is described in Ref \cite{GN} and even more completely in Ref. \cite{sciam}.  Indeed, experimenters used this parametrization up to the mid-20th century \cite{plimpton}.

Even at early times, any departure from the inverse-square law was seen to
violate an appealing geometric principle: the conservation of the number of lines of force emanating from a charge.  (The force, by definition, is
proportional to the number of lines per unit area.)   For nonzero $\alpha$, the electric flux coming out of a charge is radius-dependent -- there is no Gauss law relating charge and flux.  (See Section \ref{charge}.)

Then, just around the time of the appearance of Ref. \cite{plimpton} a
competing, scale-dependent form of deviation began to seem more appropriate.  This more sophisticated reasoning, and its development, has governed the discussion of possible deviations in later formulations of the issue.


\subsection{Photon mass from the Proca equation}
\label{proca}

The new stage arose after two break-throughs.  
The first was the electrodynamics of Maxwell and
Lorenz\footnote{
In the period 1862-1867 the Dane Ludwig Lorenz independently derived the
``Maxwell equations" of 1865, but received relatively little credit for this work \cite{kragh,jdjokun}.
}
which, when fully articulated, included among its solutions freely moving
electromagnetic waves naturally identified with light.  The second was the
realization, beginning with Einstein, that if there are particles of light that are exactly massless, as 
he inferred from Maxwell's theory with its scaling property, then 
they travel at the ultimate speed $c=1/[\epsilon_0 \mu_0]^{1/2}$.

Alexandre Proca \cite{proca0,proca1,proca2,proca3,proca4}, under the influence of de Broglie, introduced a consistent modification of Maxwell's equations which would give a nonzero mass to the photon while preserving the invariance
of electrodynamics under transformations of special relativity.
In modern notation designed to make relativistic invariance manifest, with
electric and magnetic field strengths measured in the same units, the Lagrangian density  Proca wrote is
\be
{\cal L}=-F_{\alpha\beta}F^{\alpha\beta}/4-m^2c^4A_\alpha A^\alpha/2(\hbar c)^2
\ ,
\ee
with
\be
F_{\alpha\beta}=\partial_\alpha A_\beta-\partial_\beta A_\alpha.
\ee
Here the main notational changes from Proca's original form are  to describe the photon vector potential by $A_\alpha$  and to include contravariant vectors (with the metric signature spacelike positive).\footnote{
Neither the titles nor the detailed texts of Proca's papers
indicate explicitly that this is an equation for the electromagnetic field.  Indeed, from the context it is clear that he was thinking of a charged, massive spin-1 field.   The idea that this could be identified with a massive photon came later.  We have converted to modern sign conventions for ${\cal L}$, but of course this has no effect on the free-field equations of motion.
}

This Lagrangian naturally elicits the notion that the photon might have a small but still nonzero rest mass.   The obvious implication, and the only one discussed by Proca, is a dispersion of velocity  with frequency. In fact, the classical field equations derived from Proca's start imply not only velocity dispersion, but also departures of electrostatic and magnetostatic fields from the forms given by Coulomb's law and Amp\`ere's law.  We shall see that these implications give  more sensitive ways to detect a photon mass than the observation of velocity dispersion.

The Maxwell equations as modified to Proca form are, beginning with the Gauss law,\footnote{
Here we adopt SI units.  This  is not the usual fashion in modern particle physics but it simplifies  calculation of photon mass limits from astrophysical data, which increasingly are the most pertinent sources of new and better values.
For the record, the usual notation is, in unrationalized units,
\begin{eqnarray}
\boldsymbol{\nabla \cdot}\mathbf{E} &=& 4\pi\rho-\mu^2 V,   \label{max1a} \\
\boldsymbol{\nabla \cdot}\mathbf{B} &=& 0, \\
\boldsymbol{\nabla \times}\mathbf{E} &=&
   - \frac{1}{c}\frac{\partial \mathbf{B}}{\partial t}, \\
\boldsymbol{\nabla \times}\mathbf{B} &=&
\frac{1}{c}\frac{\partial \mathbf{E}}{\partial t}
+  \frac{4\pi}{c} \mathbf{j} -\mu^2\mathbf{A},   \label{max4a}
\end{eqnarray}
}
\begin{equation}
\boldsymbol{\nabla\cdot}\mathbf{E}= \rho/\epsilon_0-\mu^2V  \ \ ,
\label{max1}
\end{equation}
where $\mu$ is the photon rest mass in units of the inverse reduced Compton wave length,
\begin{equation}
\mu = \frac{1}{\lc}= \frac{mc}{\hbar} \ ,
\end{equation}
and where $(\mathbf{A},V)$ is the now observable 4-vector potential.

Equation (\ref{max1}) implies a `Yukawa' form for the potential due to a point charge $q$ at the origin of coordinates:
\begin{equation}
V(r) = \frac{q}{4\pi\epsilon_0}
          \frac{e^{-\mu r}}{r} \ \ .  \label{gauss}
\end{equation}
Note the exponentially decreasing factor, which gives a departure from the
inverse-square law for the electric field, scaling with the length $\lc=\mu^{-1}$.

A similar phenomenon occurs in magnetism, with
\begin{equation}
\boldsymbol{{\nabla} \times} \mathbf{B}=
\mu_0\left( \mathbf{J}+\epsilon_0\frac{\partial \mathbf{E}}{\partial t}\right)
-\mu^2 \mathbf{A}   \ \  ,
\label{max4}
\end{equation}

The remaining Maxwell equations,
\begin{eqnarray}
\boldsymbol{ \nabla \cdot} \mathbf{B}&=&0 \ \ , 
\label{max2} \\
\boldsymbol{\nabla\times}\mathbf{E}&=&
    -\frac{\partial \mathbf{B}}{\partial t},  \label{max3}  
\end{eqnarray}
equivalent to the definitions of the field strengths in terms of the potentials,
\begin{eqnarray}
\mathbf{B}&=&\boldsymbol{\nabla\times} \mathbf{A} \label{pot1} \ \ ,\\
\mathbf{E}&=&-\boldsymbol{\nabla} V
-\frac{\partial \mathbf{A}}{\partial t} \ \ ,
\label{pot2}
\end{eqnarray}
are unchanged by the introduction of a photon mass.

Interestingly, although the moment Proca wrote down his equations
\cite{proca1,proca2,proca3,proca4} the finite range of static electromagnetic forces was implied, as far as we can tell Proca himself never drew this inference. (He and de Broglie \cite{broglie} focused on velocity dispersion.)  In 1935 Hideki Yukawa did recognize this consequence for a scalar or Klein-Gordon particle \cite{yukawa1}.   From the finite range of the nuclear force he predicted that a new massive particle should be found, a prediction eventually vindicated by discovery of the $\pi$ meson. In a later paper \cite{yukawa2}, Yukawa referred to Proca \cite{proca1}.  So did Kemmer \cite{kemmer}, who observed the equivalence of his 10-dimensional spin-1 solution to the 4-vector and antisymmetric tensor of Proca \cite{proca1}.

Schr\"odinger, in a number of papers \cite{scha,schA,schB,sch2},
emphasized the link between photon mass and a finite range of static forces.  Interestingly, in 1943 Schr\"odinger \cite{schB} mentioned Yukawa in this connection, but said nothing about Proca, although earlier he had mentioned de Broglie \cite{scha}.   Further, Schr\"odinger appears to have been  the first to write the two massive Maxwell equations (\ref{max1}) and (\ref{max4}) in modern format \cite{schB}. Finally, as briefly mentioned above, an  abstract but important symmetry appears to be violated by the mass terms in Proca's equations: gauge invariance.  (See Sec. \ref{gauge}.)

Two comments are in order:   First, radiation effects labeled as coming from a nonzero photon mass, including dispersion in velocity of photons (or even of classical electromagnetic waves), were found to be too small for observation.  Therefore, as already mentioned, effects in classical electrostatics and magnetostatics  
became (and today remain) the focus. 
This is despite the fact that early discussions 
seem to have been inspired by the marriage of quantum physics (implying light-particles or photons) with the special theory of relativity.

Secondly, a consequence of this proposed (Proca) deviation from Maxwell theory, like the departure from the inverse-square power law discussed earlier, is a violation of the Gauss law.  This time the term $-\mu^2V$ in Eq. (\ref{max1}) implies a density of
`pseudo-charge', compensating for the charges of ordinary electrified particles. 
As we shall discuss a little later, the pseudocharge and the ordinary charge each are locally conserved, with no ``trading" of charge density between them.  

Before continuing, this is a good point to say something about choices of scale.  Physicists constantly adjust scales to make them convenient, without needing very large or very small exponential factors.  In the case of photon mass, the limit even at mid-twentieth century was so low that all familiar choices of mass, or energy, or even frequency scale required
exponential factors.   At that time, with a lower limit on ${\lc}$ comparable with the radius of the Earth, the corresponding period of oscillation for a photon at rest, assuming that the actual rest mass saturated the limit, would have been of order 0.1 s.  At first sight that seems a very manageable number.  However, if one tries to imagine a process of producing or observing a {\it massless}  
photon of such a long period, such an experiment quickly becomes absurd.

Put differently, such low-frequency photons are essentially unobservable as single objects.  At best, one might hope to use an ultra-low-frequency circuit to detect a classical wave corresponding to an enormous number of individual photons.   Thus, in the past, 
and even more so now, the only meaningful measure of photon mass less than or equal to the limit is in terms of Compton wavelength, i.e., phenomena observable for classical, long-range, static, electric or magnetic fields.  Even though we shall quote limits on mass expressed in other terms, those values will be so far from ones we could measure directly, or ones that have been measured for any other kind of object, that they have only formal interest.  Nevertheless, for the record let us state the relations that
determine those values, using $ \hbar\equiv{\lc} mc$ :
\begin{equation}
\lc[\mathrm{m}]\equiv\frac{1.97\times 10^{-7}}{m [\mathrm { eV}]} \equiv\frac{3.52\times 10^{-43}}{m[\mathrm{kg}]},
\end{equation}
We have then in units of $c^2$,
\be
1 ~ \mathrm{kg} \equiv 5.61 \times 10^{35} ~ \mathrm{eV}.
\ee


\subsection{Conservation of electric charge}
\label{charge}

There is a deeper level in the esthetic considerations supporting vanishing photon mass, arising from an elaboration of the Gauss law.  If the electric flux out of any surface measures the total electric charge enclosed, then special relativity assures that charge must be locally conserved.  This is because the only way charge can change is by changing the flux at the same time, and for a distant surface that flux could not change instantaneously if the charge changed.   More specifically, manifest gauge invariance (ignoring the Stueckelberg-Higgs mechanism discussed in \ref{gauge}) implies local charge conservation through the invariance of the integrated
$\mathbf{J}\boldsymbol{\cdot} \mathbf{A}$ term in the action.  Thus this
conservation law is consistent with vanishing photon mass, as both follow from manifest gauge invariance.

Ogievetsky and Polubarinov \cite{ogiev1,ogiev2}, and independently Weinberg \cite{Weinberg:1964ew}, 
took a different tack.  Assuming only special relativity, they demonstrated a stronger result,  
that vanishing photon (graviton) mass by itself implies vanishing four-divergence of the electric current (energy-momentum tensor) density.  

This comes from the fact that the vector (tensor) field has only two helicity degrees of freedom, not the three (five) one would expect for a massive field.    The result tells us two things.   If the photon or graviton mass vanishes, we have a deduction of another accurately verified observation, local conservation of electric charge or of energy and momentum.      If the mass doesn't vanish, we are allowed, though not required,  to contemplate the possibility of processes violating local conservation.   

Weinberg came to this conclusion by considering the S matrix for a process involving emission or absorption of a massless photon (graviton).  Ogievtsky and Polubarinov framed their discussion in terms of field theory, and made an additional comment:  if the particle mass were finite the same conclusion of local conservation would hold provided the four-divergence of the vector (tensor) field vanished.  This condition arises naturally from considering the space-vector (tensor) wave function of a particle with spin 1 (2) in its rest frame.    There the wave function has only spacelike components and therefore obviously is orthogonal to the particle four-momentum (which in that frame is purely timelike).  This is the converse of the result going back to Proca, that his equation may only be solved consistently in the presence of a conserved current if the Lorentz gauge condition obtains.   Nevertheless, though quite a reasonable proposition, this deduction does not have the same force as that for local conservation of charge in the case of zero mass for the vector particle.

For electrodynamics 
with finite photon mass the question of electric  
charge conservation has been studied by Okun and  collaborators \cite{Okun:1978xn,Okun:1978bp, Okun:1989qi}, 
by Ignatiev and collaborators \cite{Ignatiev:1978xj, Dobroliubov:1989mr, Ignatiev:1996jr}, 
by Nussinov and collaborators \cite{Nussinov:1987jw, Mohapatra:1991as, Aharonov:1995kc},
and Tsypin \cite{Tsypin:1988bw}.  Perhaps the most interesting aspect is that coupling of longitudinal photons to electric currents, which vanishes for conserved current in the zero mass limit, now becomes divergent as the photon mass goes to zero.  In view of Ogievetsky and Polubarinov's and Weinberg's result this makes sense:  If zero mass implies conserved current, then a term violating the conservation would be ``resisted" by the electromagnetic field (i.e., its otherwise decoupled longitudinal part), which would radiate furiously to
compensate.\footnote{
If the electric current is conserved, then the $J_\mu A^\mu$ coupling for a longitudinal photon, which can be written ${ A_\mu=} \  \partial_\mu \Lambda$,
becomes zero because $\partial_\mu J^\mu =0$,
where one has used integration by parts.   However, if the current is not conserved, then this zero isn't so.  Instead, because the D'Alembertian on $A$ is $J$, one gets a radiated longitudinal $A$ field going like  $J$ divided by D'Alembertian plus $\mu^2$ in the Proca case.  In the $\mu \to 0$ limit, where the photons travel at the speed of light, this becomes divergent:  For the longitudinal part going as four-gradient of $\Lambda$ this gives rise to a $1/\mu$ 
singularity as long as the divergence of $J$ does not vanish.  
}

There is an important additional point.  Charge non-conservation destroys the renormalizability of perturbative quantum electrodynamics.  The theory begins to resemble gravity in that the latter theory also is not renormalizable (even with locally conserved energy and momentum).  Also, in the case of a massive graviton, the lowest-order perturbative theory for interaction between two gravitational sources does not limit with decreasing graviton mass to the result for zero graviton mass, as we discuss later in this article.

Thus, electrodynamics with non-conserved charge prefigures many of the features found in quantum gravity.  Nussinov \cite{Nussinov:1987jw} suggested that the regulator energy cutoff $\Lambda$ in a theory where charge conservation is violated may be connected to the photon mass by the relation $\mu\approx \delta e \Lambda$, with $\delta e$ the coefficient of an effective charge-violating coupling such as $\bar{\psi}_e\gamma_\alpha\psi_\nu A^\alpha$. 

We finally mention here an intriguing work that at least opens the possibility for a future proof that the photon mass must be identically zero.   Rosenstein and Kovner studied electrodynamics in 2+1 dimensions, and concluded that a magnetic flux condensate would form, assuring zero photon mass \cite{kovner}.  If their
method could be extended to 3+1 dimensions it might yield such a proof.


\subsection{Gauge invariance, its apparent violation 
and ultimate restoration}
\label{gauge}

Already in pre-quantum physics, the significance  of continuous symmetries such as translation invariance in space and time, as well as rotation invariance, and their links to conservation laws of momentum, energy, and angular momentum,  had been recognized.  On top of this, the Maxwell equations admit another symmetry, classical gauge invariance, or gauge invariance of the first kind.\footnote{
See the reviews of Jackson and Okun \cite{jdjokun} and of Wu and Yang \cite {wuyang} on gauge invariance, 
as well as the annotated bibliography by Cheng and Li \cite{Cheng:1988wf}, and the book by O'Raifeartaigh \cite{O'Raifeartaigh:1997ia}.  The last includes English versions of many of the key articles quoted here. 
}
From the relations (\ref{pot1}) and (\ref{pot2}) one finds that the
electromagnetic field strengths $\mathbf{E}$ and $\mathbf{B}$ are unchanged by the transformations
\begin{equation}
V\to V^\prime= V-\frac{\partial \Lambda}{\partial t} ;
~~~~\ \mathbf{A}\to \mathbf{A}^\prime =
       \mathbf{A}+\boldsymbol{\nabla} \Lambda \ \ . \label{gt1}
\end{equation}

Without the explicit $\mu^2$ terms added to two of the Maxwell equations,
Eqs. (\ref{max1}) and (\ref{max4}),  
they and the corresponding action also 
are invariant under the
transformations (\ref{gt1}).   With the  $\mu^2$ terms, and if one also assumes that the electric charge and current densities obey the equation of continuity (also known as local charge conservation)
\begin{equation}
\boldsymbol{\nabla\cdot}\mathbf{J}+\partial \rho/\partial t=0 \ \ ,
\end{equation}
one finds that the potentials must obey a restrictive condition.  This condition yields what is known as the Lorentz gauge,
\begin{equation}
\partial_\mu A^\mu = \boldsymbol{\nabla\cdot}\mathbf{A}+\epsilon_0\mu_0
\frac{\partial V}{\partial t}=
    \boldsymbol{\nabla\cdot}\mathbf{A}+\frac{1}{c^2}
\frac{\partial V}{\partial t} = 0 .
\end{equation}  
This Lorentz gauge condition is the formal expression of the fact mentioned  in Sec. \ref{proca} that consistency of the theory requires that  if ordinary charge is locally conserved, then so is the pseudocharge whose density is $-\epsilon_0\mu^2V$. 

Thus, gauge invariance appears to be broken by introduction of a photon mass.  The only allowed residual gauge transformations entail solely functions obeying the wave equation
\be
[\nabla^2-(1/c^2)(\partial/\partial t)^2]\Lambda=0.
\ee
Indeed, both for the photon (and the graviton) there was long a feeling that gauge invariance (and its gravitational analogue general coordinate invariance) provides a fundamental basis for assuming exactly zero mass.

To examine this issue more fully, we need to remind ourselves of how the form taken by gauge invariance in the context of quantum mechanics came to be.  Weyl introduced the term ``gauge invariance" in 1918$-\! 19$ \cite{weyl1,weyl2,weyl}, before the appearance of modern quantum mechanics.\footnote{
In his first two papers \cite{weyl1,weyl2} Weyl used {\it masstabinvarianz}, which means magnification or scale invariance.  It was only in his third paper \cite{weyl} that he used {\it eichinvarianz}.  A translation to German for gauge, in the sense of calibration, is {\it eichmass}.   (A train track gauge is {\it spurweite}, totally different.)
}

He wanted the gravitational metric and the electromagnetic field to transform as
\bea
g_{\mu\nu}(x) &\rightarrow& e^{2\alpha(x)}g_{\mu\nu}(x),
       \label{Wconformal} \\
A_\mu(x) &\rightarrow& A_\mu(x) - e\partial_\mu \alpha(x).
\eea
(Here gauge is used in the sense of scale, because $\alpha$ is real.)  This type of change is now known  as a conformal or scale transformation.

An early paper of Schr\"odinger \cite{schweyl}, written before the wave equation was discovered, 
speculated about the possibility that for an allowed closed particle path the loop integral of the vector potential might be quantized. 
Then shortly after the appearance of 
wave mechanics, 
Klein \cite{OK}, Fock  \cite{fock}, 
and Kudar \cite{kudar}, each 
emphasized (in the context of a five-dimensional formulation linking electrodynamics to mechanics) 
the pre-requisite idea that the full electromagnetic interaction in the relativistic form of the Schr\"odinger equation 
(today known as the Klein-Gordon equation) 
entails the form $-i\hbar\nabla-q\mathbf{A}$.  This was later used by Weyl (see Eq. (\ref{fullS.E.} below).

Schr\"odinger himself \cite{schim}, Gordon \cite{wgordon},  and London
\cite{flondon,flondon2} made the same point, also in the context of the relativistic equation.  London \cite{flondon2} connected the discussion to  Weyl's pre-quantum work as well as Schr\"odinger's early paper \cite{schweyl}.

For Weyl, gauge invariance was a ``master principle" to govern construction of the theory,  
and in 1929 he 
revised his approach for electromagnetism in quantum mechanics  \cite{weylA,weylB,weylC}, setting the stage for all future discussions.  Instead of his first idea of a scale transformation, he introduced a phase transformation of the wave function.  He considered the Schr\"odinger equation for a particle with electric charge $q$
\begin{equation}
i\hbar\partial_t\psi =\left[\frac{(-i\hbar\boldsymbol{\nabla}-q\mathbf{A})^2}{2m}
+qV \right]\psi \ \ . \label{fullS.E.}
\end{equation}
Under the simultaneous transformations (\ref{gt1}) and
\begin{equation}
\psi\to \psi^\prime =e^{iq(\Lambda/\hbar)}\psi   \ \ ,
\end{equation}
we see that the Schr\"odinger equation is unchanged.  This is known as gauge invariance of the second kind.\footnote{
Fock both in the title and the content of his paper \cite{fock} had described and exhibited gauge invariance of the second kind without the name, and without commenting on its significance.)
} 

Two decades after Proca introduced his mass mechanism enforcing the Lorentz gauge, Stueckelberg found what initially may have seemed merely a formal way of keeping mass and at the same time restoring gauge invariance \cite{ecgS}.  He introduced
a new scalar field, $\Phi$, with fixed magnitude and carrying electric charge, $q$, whose `kinetic', gauge-invariant contribution to the Lagrangian density is
\begin{equation}
{\cal L}_{\rm S}=\frac12 \left[\left|
-\partial_t\Phi+iqV\Phi/\hbar\right|^2
-\left|\boldsymbol{\nabla}\Phi-iq\mathbf{A}\Phi/\hbar\right|^2\right]  \ \
.
\label{stueck}
\end{equation}

Here we are dealing with a Klein-Gordon equation, rather than a
Schr\" odinger equation.  Otherwise this is simply an example of the new gauge invariance required in quantum mechanics, even though we
may treat $\Phi$ as a classical field.  At this point we may choose a gauge  by assuming that the phase of $\Phi$ is zero everywhere or, indeed, has any constant value.   In that gauge, it is easy to see that the
extra term in the action becomes
\begin{eqnarray}
{\cal L}_{\rm S}&=&-\frac12 \mu^2({\mathbf{A}}^2-V^2) \  , \\
\mu &\equiv& q\Phi/\sqrt\epsilon_0\hbar c
=q\Phi\sqrt\mu_0/\hbar \ .
\end{eqnarray}
This is just the Proca photon mass term we have seen before, again with mass expressed in units of inverse (reduced) Compton wavelength.  Now, however, the restriction to Lorentz gauge comes only because we made a specific choice (zero) for the phase variation of $\Phi$.  With no such specification, full gauge invariance is restored, even though the photon now has a non-zero mass.

Therefore we may replace the earlier guess, that gauge invariance implies zero photon mass, by a new, more precise assertion:  The minimal dynamics obeying gauge invariance  (the Maxwell action) implies zero photon mass.  However,  by adding more dynamics, for example,  another field $\Phi$ interacting with the photon field, we may keep gauge invariance and accommodate non-zero mass at the same time.

If we think of variation in  the phase as a (spacetime) position-dependent
rotation, then it is immediately clear that the corresponding symmetry must be unbreakable as well as unobservable:   Observable arbitrary position-dependent rotations would put arbitrarily great stresses on any system, and thus could not be symmetries.

For reasons like this, gauge invariance and general coordinate invariance have (sometimes) been called ``fake" symmetries.   This term should be treated with care, since it could be taken to imply that the whole concept is useless.  However, as we have seen, this abstract and unobservable symmetry, infinitely flexible and therefore intrinsically unbreakable, provides a powerful organizing principle for dynamics.  It has an especially simple and esthetic starting point, the minimal theory, namely Maxwell theory, for electrodynamics (and of course general relativity for gravity).\footnote{
In condensed-matter physics, the compatibility of non-zero photon mass with gauge invariance is well-known.   The simplest example is a plasma, where plasma longitudinal and transverse sound waves combine to provide the three degrees of freedom one expects for a massive spin-1 particle.   Of course the plasma fixes a local rest frame, so that Lorentz invariance is broken.   An insulator admits electromagnetic excitations of arbitrarily low energy, which might make them seem massless, yet the excitations travel at subluminal speed compared to light
in vacuum.
}

For the Abelian theory, i.e., electrodynamics with a massive photon,  the Stueckelberg formulation restores explicit gauge invariance under very general assumptions about the dependence of the Lagrangian on the four-vector potential $A_\alpha$.   One simply replaces any $A_\alpha$ by $A_\alpha - \partial_\alpha \chi/\mu$, where $\chi$ is a scalar field which transforms under a gauge transformation by $\chi\to \chi + \mu\Lambda$. This linear addition is an adaptation of the approach described in Eq.$ \ $(\ref{stueck}),  involving the gauge-covariant derivative $D_\alpha \Phi=(\partial_\alpha-ieA_\alpha)\Phi$.

Section \ref{charge} suggests a complementary viewpoint to the one emphasizing gauge invariance:   Here we get the physical consequence of current conservation from physical symmetries in special relativity, and discover that there are only two helicity degrees of freedom for the massless photon and graviton fields.  Gauge invariance directly eliminates one degree of freedom from the four-vector potential.  Indirectly, by enforcing the square of the Maxwell field tensor as the kinetic term in the Lagrangian, gauge invariance forbids 
a conjugate momentum for the time component $A_0$, which therefore is not an independent variable.  Again the original four degrees of freedom  are reduced to two.  

Thus,
gauge invariance and Lorentz invariance for couplings of massless vector or tensor particles become equivalent in all respects.   Explicit application of both these symmetries provides a powerful guide both for calculation and also for insight into the structure of a theory.  For example, checking perturbative calculations for gauge invariance is a common technique to validate the calculations. Of course such calculations always are formulated in a manner guaranteeing Lorentz invariance, not only for coupling of the vector or tensor field.  


\subsection{Higgs mechanism, or hidden gauge invariance}
\label{higgs}

Though Stueckelberg's construction \cite{ecgS} removed the formal gauge-invariance argument for zero photon mass, there still was little motive for going beyond the minimal theory.  The physical interest in doing so began with the work of Yang and Mills \cite{ym}, who proposed the idea of a more elaborate gauge symmetry, where the rotations are in a three-dimensional space, rather than the single phase or
rotation angle (corresponding to a two-dimensional space) found in
electrodynamics. In later years their proposal was generalized by many authors, leading to the conclusion that gauge symmetries can apply for arbitrary compact transformation groups.  The immediate question arising when one contemplates such 
non-Abelian 
gauge symmetries is, ``Where are the corresponding massless
photon-like particles?"

One answer to this question had its intellectual beginnings in the early 1960s with the work of a number of authors \cite{schw}-\cite{higgs3}.   Schwinger \cite{schw,schw2}, and more explicitly Anderson \cite{anderson}, followed by  Englert and Brout \cite{EBrout}, by Higgs \cite{higgs1,higgs2,higgs3}, and by
Guralnik, Hagen, and Kibble \cite{GKib}, found increasingly clear ways to describe gauge particles possessing mass while the underlying gauge invariance remains unbroken.\footnote{
Possibly the first discussion of the non-Abelian version of the Higgs mechanism was in a remarkable paper representing an independent discovery of the mechanism by Migdal and Polyakov \cite{Migdal:1966tq}.
}

The big change from Stueckelberg's idea, in what has become known as the Higgs mechanism, is to allow the magnitude as well as the phase orientation of the `mass-generating' field to become dynamical.  In its simplest form, this corresponds to adding a term to the Lagrangian density of Eq. (\ref{stueck}), yielding
\begin{eqnarray}
{\cal L}_{\rm S}&=&\frac12 \left[\left|
-\partial_t\Phi+iqV\Phi/\hbar\right|^2
-\left|\boldsymbol{\nabla}\Phi-iq\mathbf{A}\Phi/\hbar\right|^2\right]
       \nonumber \\
&~& ~~~ -\frac\lambda4(|\Phi|^2-v^2)^2 \ \ ,
\end{eqnarray}
where $v$ is called the ``vev" or ``vacuum expectation value" of the field
$\Phi$: $v \equiv \langle |\Phi| \rangle$.

This ``Higgs mechanism" is a relativistically invariant analogue of the behavior of a superconductor, where a collective wave function of many charged particles  leads to damping of electric and magnetic
fields.\footnote{
In a superconductor the macroscopic electron-pair condensate wave function produces an effect like that of a non-zero photon mass, again without breaking gauge invariance.
}
The simplest form of this mechanism introduces a charged scalar field which in the ground state of the system has nonzero magnitude everywhere.

Varying the action with respect to the four-vector potential, $A^\mu=(V , {\bf A})$, yields exponential damping of a static, electromagnetic field in space and so, of course, a dispersion (small though it might be!) of wave or photon velocity  with frequency.  This corresponds to the introduction of a finite mass for the gauge boson (the photon).  As in the superconductor case, a sufficiently strong  electromagnetic field, or sufficiently high temperature, can force $\Phi$ to vanish in some region (which was not possible for the Stueckelberg field), in which case the photon may exhibit zero effective mass in that region.

Despite their appeal, these ideas lay dormant for nearly a decade, until 't Hooft's proof \cite{hooftA,hooft} that such a theory fits into the pattern established by quantum electrodynamics, a renormalizable perturbative quantum field theory.   This means that, for phenomena where the gauge coupling can be treated as small, there is a well-defined, systematic expansion in powers of the
coupling (depending only on a finite number of parameters)  to deduce
precise values for cross sections and other observable 
quantities.\footnote{
't Hooft's result required the full Higgs mechanism, in which the magnitude of the Higgs field can vary.  Only for the Abelian case, as for a massive photon, does perturbative renormalizability obtain for a Stueckelberg field of fixed magnitude.
}  

The Higgs mechanism became an integral part of the highly successful standard model unifying weak and electromagnetic interactions.  It should be noted that a more complicated form of this mechanism, in which (as is true for superconductivity) there is no particle corresponding to quantum fluctuations of the Higgs field, remains a logical possibility.  Even more important than a
possible Higgs particle as a validation of this view of electroweak interactions is the theoretically predicted and experimentally confirmed existence of the massive gauge bosons ${\rm W}^\pm$ and ${\rm Z}$.
Another non-Abelian theory, quantum chromodynamics, though with a different mechanism (color confinement) for avoiding free massless gauge bosons, has been similarly successful in describing the strong interactions.

Surprisingly, until recently the option of using the Higgs mechanism to
parametrize possible deviations from Maxwell theory remained relatively
unexplored.  Indeed, we know of only two attempts to apply these ideas to the  question of a possible photon mass.


\subsubsection{Temperature effect}
\label{temp}

Primack and Sher \cite{primack} focused on an effect familiar in
superconductivity, that above a critical temperature the condensate disappears.   Thus, they considered the possibility that at very low temperatures there might
be a Higgs mechanism that would generate a small photon mass, but at higher temperatures photons would be massless. Though they did not view this as especially likely, it still is worthwhile to examine the notion a bit  more closely.

For the condensate value $\Phi$ to disappear in a large region of space, the energy density corresponding to a given temperature in that region, $\sim (kT)^4/(\hbar c)^3$, should exceed the vacuum energy density $\frac\lambda4 v^4$  associated with vanishing $\Phi$.  This may happen either because of a very small value of $\lambda$ or a small value of $v$, or of course a combination.  Also, $\lambda$ and $v$ may be temperature-dependent, yielding a zero value for
$v$ at sufficiently high $T$.   

For the condensate to be restored in a volume characterized by length $L$, the temperature in that region must fall below a critical value.  Further, the gradient energy $\sim v^2L$ must be smaller than the vacuum energy $\sim \frac\lambda 4 v^4L^3$; that is, $\lambda v^2L^2 \ge 1$.  Thus, if the coupling $\lambda$ were too small the effect would not occur, even
if the temperature in the region were low enough.

At the same time, to find a detectable photon mass effect inside that region, one must be sensitive to a term of order
\be
(\mu L)^2=\frac{(q v L)^2}{(\hbar c)^2 \epsilon_0}=(q v L/\hbar)^2\mu_0.
\ee
This illuminates the difficulty of
implementing this mechanism:  If $q$ were appreciable but the effective mass of the scalar field were small, then one would expect to observe production of light charged ``Higgs" particles, which have not been seen.  Thus $q$, the charge of the Higgs field, must be quite small,  yet at the same time the Higgs mass must be small\footnote{
As pointed out in Section \ref{charge}, limits on the magnitude of possible small electric charges carried by very light particles have been discussed before \cite{Mohapatra:1991as}.  Also see \cite{Davidson:2000hf}.
}
and the charge $q$ sufficiently large, so that the Primack-Sher temperature effect would be observable.  This leaves at best a very small region in the three-dimensional parameter space $(q,v,\lambda)$ for which the effect would be possible \cite{abbott,preply}.


\subsubsection{Large-scale magnetic fields and the photon mass}
\label{largemag}

Recently Adelberger, Dvali, and Gruzinov (ADG) \cite{adel} proposed using this type of mechanism to parametrize possible deviations from Maxwell theory on large length scales.  Like Primack and Sher they used an Abelian Higgs field (not related to the standard-model Higgs).
ADG's most striking point is that a phenomenon like that of Abrikosov vortices in a superconductor could allow a substantial mean magnetic field $\mathbf{B}$ over galactic or even larger regions.

The Higgs field would have null lines parallel to the direction of $\mathbf{B}$, while the phase of the field would circulate with period $2\pi$ about each line.

If one did not happen to be sitting near a vortex line, extremely precise local measurements of
{\it electric} fields could indicate patterns associated with a tiny nonzero photon mass.
Even so, the implication from Proca theory -- that a nonzero
average field over a region
requires an upper bound on the photon mass -- 
would no longer hold.\footnote{
The region would have dimensions transverse to the field direction characterized by a  very large length scale $L$.  
Provided the photon magnetic  Compton wavelength were large compared to the typical separation between vortex lines, $\mathbf{B}$ would be essentially constant.
}

The basis for this effect goes back to Stueckelberg's  observation \cite{ecgS}, discussed above.  By transforming to a gauge in which the Higgs field has constant phase, one obtains a vector potential
\begin{equation}
\mathbf{A}=\mathbf{A}_{\rm Maxwell}- \boldsymbol{\nabla}  \Lambda \ \ .
\end{equation}
Then the energy density contribution
\be
{\cal E}_{\rm photon-mass}=|q^2\Phi|^2 \mathbf{A}^2/2
\ee
is suppressed compared to the Proca case because the phase vortices make the average  $\! \langle\mathbf{A}\rangle$ vanish. In fact the average
``photon-mass" energy density is reduced to
\be
\langle {\cal E} \rangle ={\cal O}(\mu^2 \ell^2)B^2/\mu_0
\ee
instead of ${\cal O}(\mu^2L^2)B^2/\mu_0$, where $\ell$ gives the typical
separation between vortex lines and $L$, as before, is the typical length
transverse to $\mathbf{B}$ over which $\mathbf{B}$ is roughly uniform.

For fixed $\lambda$ and $v$, and $B<\sqrt{\lambda \mu_0}v^2$, as $\mu$ decreases ($\lc$ increases) the Higgs field becomes increasingly `stiff', and this more complicated theory reduces to the second-stage or Proca form.  The possibility of achieving such a limit demonstrates that there exists a mathematical transformation which formally restores gauge invariance to the Proca theory, just as had been observed by Stueckelberg \cite{ecgS}.  Thus,  previous limits on the Proca mass remain valid provided $\lc$ is not too small, so that there is a smooth continuity between stages 2 and 3.

A novelty of the more general (Higgs) form, in addition to possible measurements of apparent photon mass, is that in the regime of moderate or small $\lc$ one also has possible observations of critical field or critical temperature effects associated with extinction of the mass.  In particular, one may consider a regime where the typical field strength $\mathbf{B}$ is so great that $\langle \Phi \rangle$ is brought to zero.   Now one has a situation quite similar to that discussed by Primack and Sher for temperature \cite{primack}, where much of space shows no photon mass. Still,  in a sufficiently large region of true vacuum, with sufficiently small $\mathbf{B}$ or $T$, one possibly could detect a nonzero and perhaps even quite substantial mass, for example by repeating the WFH experiment there. (See Sec. \ref{WFHexpt} below and Ref. \cite{wfh}.)


\subsubsection{Empirical and formal considerations on the Abelian Higgs
mechanism}

From the viewpoint of testing this Abelian Higgs
concept, there is a major change from the fixed Proca mass.
This time there are three parameters, (i) the optimum, energy-minimizing
magnitude of the Higgs field vev, $v$, (ii) a coefficient of assumed quartic self-coupling of the Higgs field, $\lambda$, and (iii) a parameter, $q$, representing the charge of the Higgs field which determines its coupling to the electromagnetic field.
This increases the challenge of determining the parameters, or even limits on them.

At the same time this gives more observational tools for constraining the
parameters.   For example, if the particles had low mass, then their charge would have to be very small, because otherwise they would be created copiously, and easily detected, in any high-energy process involving collisions of ordinary charged particles.  Clearly there is incentive for  followup work, beyond the discussion for the ``zero-temperature" case presented by ADG, to map out regions in the three-dimensional parameter space still allowed by existing measurements.  This also could determine what further measurements might best improve the constraints on the allowed parameter domains.

While the massive gauge bosons of electroweak interactions show that
gauge-invariant mass of gauge particles is possible, there may still be
constraints of principle.   First, extensive studies of self-coupled scalar fields indicate that such a system would
only make sense if the dimensionless coupling $\lambda\hbar c$ were of order unity or less.  Secondly, the dimensionless gauge coupling in  electroweak interactions is comparable for the electric and the weak sectors.  This makes the domain of possibilities opened up by the discussions in 
Sections \ref{temp} and \ref{largemag} seem questionable, because they inevitably would entail an exponentially smaller electric charge for the Abelian Higgs field than for any other particle.\footnote{That small charge has as a possible consequence that
the time for the field to come into equilibrium with a nonzero value at very low temperature would be too long for practical observation.}
Meanwhile, the limitation on $\lambda$ excludes a strictly  fixed photon mass (although for large enough $v$ the Proca-Stueckelberg limit could be an excellent approximation).

Thus, modern quantum field theory gives some arguments to suggest that  there may be no photon mass at all (even of the ``gauge-invariant" type), reinforcing older considerations such as the geometrical significance of the Gauss law and the appeal of the minimal gauge coupling hypothesis seen in Maxwell theory.  While the Gauss law relating charge to electric flux is broken explicitly for the deviation from Coulomb's law considered in stage one, it can be argued that it still holds  formally for  stage two and physically for stage three.  This is
because the vector potential in stage two and the electrically charged Higgs field in stage three can be taken to contribute to the electromagnetic charge and current densities.

Thus, if one looks at things in a certain way the symmetries and conservation laws apparently broken if a photon mass effect were observed could be said merely to be hidden.
In any case, despite the lack of positive observations up to now, the issue of a nonzero mass of course remains open, because an exact zero can never be established by experiment.

The evolution we have described entails increasing numbers of parameters for an assumed deviation of classical electrodynamics from a strictly Maxwellian form.  At the same time, there are more phenomena which can be examined to test for the deviation.  Thus, the process of testing becomes more demanding, but the accuracy of limits in principle can be maintained or possibly improved.


\subsection{Zero-mass limit and sterile longitudinal photons}
\label{phlimit}

There is a profound conceptual discontinuity associated with the zero-mass limit of massive electrodynamics.  For any nonzero mass, there are three degrees of freedom, corresponding to the three possible orthogonal polarizations of a photon in its rest frame.  Nevertheless, all observable phenomena of electrodynamics are continuous in the limit.
Part of the reason is that as long as electric charge is locally conserved the coupling amplitude for radiation of longitudinal photons is suppressed by a factor ${\cal O}( \mu^2/k^2)$ for photons of wavenumber $k$.  Thus for any fixed $k$ the coupling vanishes as $\mu\to 0$.  In the limit then, longitudinal photons exist, but are completely invisible, or ``sterile."

An important qualification:  The above statement about the near-sterility of the longitudinal photon at very low mass need not apply when gravitational interactions are taken into account.  For example, the longitudinal component should contribute equally with transverse components to the energy-momentum tensor.  One might ask, then, if a longitudinal photon could be a viable candidate for a possible ``dark-matter" particle, as discussed elsewhere in this paper.  The answer seems to be negative, because cold  dark matter would be hard to account for if the mass were so far below any plausible scale for the temperature of these particles.  

Ogievetsky and Polubarinov \cite{opnotoph} gave an instructive analysis relevant to these issues.   They pointed out that in the $\mu=0$ limit of Proca theory the helicity-zero or longitudinal state disappears from the kinetic energy (and therefore would not contribute to gravitational couplings).  Nonetheless, if one starts with Kemmer's tensor formalism for massive photons \cite{nk2}, in the zero-mass limit {\it only} the zero-helicity state survives.  This difference holds even though the two formalisms are equivalent for any nonzero mass.\footnote{
Thus both formalisms constitute examples of a discontinuity at zero mass, a phenomenon seen even more dramatically in linearized gravity as discussed in Sec.\ref{vdv}.
}   
They call the zero-helicity state the ``notoph."\footnote{
The meaning is discernible in the Latin alphabet, but even easier in the Cyrillic.
} 
Further, the notoph does not couple to any conserved current,  so (in the discontinuous zero mass limit) it {\it only} should be detectable gravitationally.  

Bass \cite{bass} examined 
a possible explanation for cooling of the Earth's core, by emission of slightly coupled longitudinal photons.   However, using Schr\"odinger's and Schr\"odinger and Bass's earlier estimates \cite{schB},\cite{sch2} of a limit on photon mass from the properties of the Earth's magnetic field, he could rule  out cooling by longitudinal-photon emission -- the coupling is far too weak.

Even for static or low-frequency phenomena, the relative deviations of
electromagnetic fields from their values for  zero $\mu$ are  small, ${\cal O}(\mu^2L^2)$, where $L$ is a characteristic spatial dimension of the region under study \cite{GN}.   Although one is not looking at radiation here, the root cause for the suppression factor is the same.  This can be understood by asking what would be the typical wavenumbers of virtual photons associated with such a configuration.

As mentioned in \ref{charge}, if electric charge were not locally conserved then longitudinal photons would be super-strongly coupled.   Thus the continuity of the zero-mass limit depends on delicate cancellations that could easily be upset.   Nevertheless,
as long as local charge conservation holds, the continuity 
of elecrodyanamics 
applies not only for observable electromagnetic fields in vacuum but also for fields in all kinds of material backgrounds.

Interesting examples of this statement include:  1) The
continuity of the index of refraction and other electromagnetic quantities in $\mu$ implies that the recently discovered phenomena of "fast" and "slow" light \cite{milonni} should not be affected  by  a small Proca mass.  2)  The same applies even to explicitly quantum phenomena, such as the well-known Casimir effect of attraction between two uncharged conducting plates \cite{Barton:1984icA,Barton:1984icB}.


\section{Secure and speculative photon mass limits}
\label{EMtests}

Quoted photon mass limits have at times been overly optimistic in the strengths of their characterizations.  This is
perhaps due to the temptation to assert too strongly something one ``knows" to be true,  A look at the summary of the Particle Data Group \cite{pdg} hints at this.  In such a spirit, we here give our understanding of both secure and speculative mass limits.

The key to intuitively understanding the new physics is to solve the
time-independent Proca equations (\ref{max1})-(\ref{max4}).  In particular the electric potential is not the Coulomb potential but a Yukawa potential.  Putting Eq. (\ref{gauss}) in simplified form, it is
\be
V(r) = - \frac{e}{r} \exp[-\mu r],  \label{yukawa}
\ee
where again $\mu$ is the mass expressed as the inverse (reduced) Compton wavelength. A similar Yukawa fall-off occurs for the magnetic vector potential and field.   By taking the gradient of Eq. (\ref{yukawa}) one finds that the first non-Coulombic term in ${\bf E}$ is of order $(\mu r)^2$.  This size turns out to be general, and can be given by a theorem \cite{gnth}.

Therefore, as we \cite{gnth} and others \cite{parkth,kroth} have emphasized, to measure a small photon mass you need either a very precise experiment or a very large apparatus.  That is, a precise experiment can measure the very small deviation from unity in a slowly falling exponential and a very large apparatus has the advantage of having a large exponential fall-off vs. unity.
Since the publication of Ref. \cite{GN} there have been  extensions of
previously introduced approaches to do this,  and also 
three new ideas.


\subsection{Local experiments}


\subsubsection{Electric (``Cavendish") experiment}
\label{WFHexpt}

Laboratory tests of Coulomb's Law are the cleanest one can perform.  This is not surprising, as the experiments are small and local.  They can be repeated, and systematic uncertainties can be characterized and reduced, obviously important here.  Since the apparatus is ``small" a precise experiment is necessary.  It is both a tribute to their ingenuity and also a comment on how the size of an experiment limits a photon mass measurement, that the 35-year-old result of Williams, Faller, and Hill \cite{wfh} remains the landmark test of Coulomb's Law.  Their limit of
\bea
\lc &\gtrsim&  2\times 10^ 7 \,\mathrm{m}, \ ~~~ \mathrm{or}  \nonumber \\
\mu &\lesssim&  10^{-14}\,\mathrm{eV}
\equiv 2 \times 10^{-50}\,\mathrm{kg}  \ ,
\eea
is unsurpassed in the substantiated (laboratory)
literature.\footnote{
A later reanalysis proposed a smaller number \cite{wfh+}.  Around the same period, a small improvement was claimed in \cite{crandall}, but the result was never published to our knowledge.
}


\subsubsection{Magnetic (Aharonov-Bohm) experiment}

Boulare and Deser \cite{Boulware:1989up} observed that another null experiment can be done with a toroidal 
magnetic field confined by a 
superconducting ``skin." 
The flux inside the superconductor must be an integer number of flux quanta, but with nonzero photon mass there will be an antiparallel flux outside in the vicinity of the superconductor suppressed by a factor of
${\cal O} (\mu^2 \ell^2)$, where $\ell$
is a characteristic dimension of the apparatus.   They estimated that an
experiment of this sort could produce a limit
$\lc\gtrsim  10^ 5  \mathrm{m}$.  To the best of our knowledge no such dedicated experiment yet has been performed.  We suspect that with the help of a SQUID detector their proposed sensitivity could be improved, but perhaps not to the level of the result in \cite{wfh}.


\subsubsection{Temperature effect}

The  ideas of Primack and Sher \cite{primack} on a photon phase transition at low temperature, even if incomplete \cite{abbott,preply}, inspired a
low-precision ($\lc\gtrsim 300$ m) experiment.by Ryan, Acceta, and Austin \cite{ryan}, performed at 1.36 K.\footnote{
An experiment considered by Clark was never completed to our
knowledge.  See Ref. \cite{dombeyexpt}.  Such discussions also stimulated the late Henry Hill, who expressed strong interest in performing a Coulomb's Law test at very low temperatures (mK range)
to search for a phase transition.  (See Ref. \cite{mmncline}.)
}
As we have mentioned already, this negative result need not be meaningful, because (1) gradient energy of the Higgs field could prevent its acquiring a nonzero value in a small region maintained at low temperature, and (2) a very small electric charge of the field could keep it from coming into thermal equilibrium during a time practical for observation.


\subsubsection{Dispersion, radio waves, and the Kroll effect}
\label{Kroll}

For decades de Broglie hoped to find a photon mass, at first by the dispersion of optical light from stars.  He performed a calculation in 1940 that claimed a limit of $\mu \le 10^{-47}$ kg  \cite{broglie}.
However there was a numerical error of $\approx 10^5$ \cite{GN}, which when corrected yields
\bea
\lc &\gtrsim&  0.5 ~\mathrm{m}, \ ~~~ \mathrm{or}  \nonumber \\
\mu &\lesssim& 4 \times 10^{-7}\,\mathrm{eV}
\equiv 0.8 \times 10^{-42}\,\mathrm{kg}  \ ,
\eea

In our article \cite{GN} we discussed at length the dispersion in pulsar waves, which is an easily measurable effect.   However, this dispersion is commonly accepted as a measure of the density of interstellar plasma.   Interpreted as a photon mass it would give a value far above that excluded even by laboratory experiments \cite{feinberg}.  Because pulsar signals have such a long flight path, we incorrectly assumed that no better result could be found from velocity dispersion.

At about the same time as  our previous review appeared, Kroll \cite{Kroll:1971wi} discovered a way to do something we had thought impossible -- obtain a reasonably competitive limit on photon mass from wave velocity dispersion, He did this by using Schumann resonances, which
are very low-frequency standing electromagnetic waves traveling through the atmosphere parallel to and between the Earth's surface and the ionosphere, two conductive layers.

There are two important considerations here.
First note that for a wave propagating  between and parallel to two plane conducting layers, perhaps surprisingly there is a special mode whose speed is $c$, {\it even} if there be a non-zero photon mass \cite{GN}.
Two concentric spherical conducting layers (with slightly different radii $R_>$ and $R_<$) of course are not exactly parallel to each other.  Even so, Kroll found that now the mass contributes to velocity dispersion of the special mode, but with $\mu^2_{eff} = g\mu^2$.   Here the dilution factor $g$ for the modes that would travel at speed $c$ between parallel conductors is of order $({R_>-R_<})/({R_>+R_<})$, which in this case would be slightly less than 1\%.
This means that the limit obtained on the photon mass would be only an order of magnitude worse than naive expectations (i.e., expectations in ignorance of the behavior of the special mode) might have suggested.

The second point is that the atmosphere between the two conducting layers has a conductivity far smaller than that of the interstellar plasma.\footnote{
The mobile electron density in the atmosphere is much higher than in interstellar plasma.  However, electron-atom collisions quench the electron contribution to atmospheric conductivity, and the dominant contribution to conductivity from ions still is small.
}

Thus, by looking at really low frequencies (where the lowest is about 8 Hz), one may obtain an interesting limit even for waves whose travel distance is no more than the circumference of the Earth.
Kroll deduced a limit
\bea
\lc &\gtrsim& 8\times 10^5\,\mathrm{m}, \ ~~~ \mathrm{or}  \nonumber \\
\mu &\lesssim& 3 \times
10^{-13}\,\mathrm{eV} \equiv 4 \times 10^{-49}\,\mathrm{kg}  \ ,
\eea
i.e., $\lc$  about a tenth  the radius of the Earth.

Recently F\"ullekrug \cite{fullekrug} adapted Kroll's method to new and more refined data on the Schumann resonances and the height of the ionosphere.   He claimed a result about three orders of magnitude better than Kroll's.  F\"ullekrug made the assumption that the frequency shift due to photon mass $\mu$ is linear rather than quadratic in $\mu$.    His assumption is contrary to the theorem \cite{gnth,parkth,kroth} discussed in the preamble of this section, and therefore leads us to strong reservations about the details of his approach.

A possible explanation for his assumption is that according to his analysis a fractional shift in  circular frequency $\omega$ {\it is} equal to the ratio   $A= (\Delta h_2)/(2\sqrt{h_1h_2})$, where $h_2$ is the ionosphere height (about 100 km), $\Delta h_2$ is its possible fluctuation, and $h_1$ is the height of that point in the atmosphere where the displacement current and the electric current are equal in magnitude (about 50 km).    Simply from dimensional analysis, he likely is right that this effect on wave phase velocity is linear in the quoted ratio.   However, because the Maxwell equations involve $\mu^2$  we do not see how there can be a linear dependence of phase velocity on a very small photon mass.

In our view the proper way to obtain an optimum limit on photon mass from these data would be to fit deviations in the  lowest frequencies to the formula
\be   
 \delta\omega_i=A\omega_i+B/\omega_i \ \  ,
\ee   
with $B=g\mu^2c^2/2$.  Unfortunately the data presented in the paper are insufficient to carry out this fit.   We think that, although it is likely there would be a significant improvement over Kroll's result,  it would not be by three orders of magnitude.


\subsection{Solar system tests}

\subsubsection{Planetary magnetic fields}

The idea of Schr\"odinger to test for a photon mass by measuring the Earth's magnetic field \cite{schB,sch2} took advantage of the other side of the laboratory paradigm with its precise measurements.  Instead one uses a large,  though less refined, apparatus.
Over the years a number of improvements were made to Schr\"odinger's method for the Earth \cite{GNearth,ephearth}.

The best current result of this type came from using an even bigger apparatus, Jupiter.  A limit of
\bea
\lc &\gtrsim&  5\times 10^8\,\mathrm{m}, \ ~~~ \mathrm{or}  \nonumber \\
\mu &\lesssim& 4 \times
10^{-16}\,\mathrm{eV} \equiv 7 \times 10^{-52}\,\mathrm{kg}
\eea
came from the Pioneer 10 flyby of Jupiter \cite{DGN}.
We emphasize that because this limit is due to data from the first flyby of Jupiter, it was calculated in an extremely conservative manner, at least by a factor of 2.  Furthermore, with modern data a more precise number could be obtained.

However, once again, because of the $(\mu r)^2$ effect, an order of magnitude improvement basically calls for an order of magnitude larger magnet, say the Sun.  Ideas on how a solar probe mission could do this were given in Ref. \cite{KN}.


\subsubsection{Solar wind}

Finally, there is the (geometrically) largest magnetic field in the solar system, that
associated with the solar wind.  In principle this could yield the best directly measured limit.  Using the MHD equations for a finite Proca mass
and a generous upper bound for the $\mu^2A^2$ energy of the solar wind magnetic field, Ryutov \cite{ryutov1}  found some time ago that a limit at ``a factor of a few better" than the Jupiter limit should follow.

Recently Ryutov has been able to use fuller data on the plasma and magnetic field, extending to the edge of  the solar system, to make a  dramatic further improvement \cite{ryutov2}:
\bea
\lc &\gtrsim& 2\times 10^{11} ~{\mathrm{m}} ~~~~
            {\mathrm{or}} \nonumber \\
\mu &\lesssim& 10^{-18}~ {\mathrm{eV}} \ \equiv 2\times                              10^{-54}~{\mathrm{kg}} ,
\eea
or a minimum reduced Compton wavelength about 1.3 AU.

To understand Ryutov's method one needs to know something about the expected (and found!) form of the magnetic field associated with the solar wind.   Parker \cite{parker} worked this out long before the distant satellite measurements:   The radially moving plasma carries with it the magnetic field lines, and (because the Sun rotates) these field lines "wind up" like an Archimedes spiral.  Thus, at large distances the field is principally azimuthal.

To maintain this field if there were a photon mass,  there would have to be an actual current to cancel the Proca current $-\mu^2\mathbf{A}/\mu_0$ implied by the Proca equations.   The satellite observations do not measure current directly, but they do give plasma density, plasma velocity, and plasma pressure at least out to the orbit of Pluto.  If there were such a current, then the resulting ${\bf J} \times {\bf B}$ force density would cause a calculable acceleration of the plasma.   The data limit any such (both radial and polar-angle) acceleration, thus providing an upper limit on $\mu$.

To account for the partial angular coverage of the satellite observations, Ryutov allowed an extra order of magnitude in the limit on ${\bf J}$, and hence a``safety factor" of three  in the limit on $\mu$.  Thus he obtained  $\lc\ge 1.3 $ AU as a clearly conservative limit.  To the best of our knowledge, this is the strongest limit in the research literature supported by well controlled and understood data.


\subsection{Cosmic tests}

With the new solar wind results \cite{ryutov2} we may have arrived at the end of the era in which direct ``local" laboratory experiments could contribute to limits on $\mu_\gamma$.   These results are based on experiments using apparatus in satellites to measure magnetic fields and plasma currents in large parts of the solar system.   It is hard to imagine how such direct observations could be carried to much larger distances.   Thus, further research must rely on observations of radiation from more remote regions, as well, possibly, as observation of $\mu^2{\mathbf{A}}$.


\subsubsection{Fields on galactic scales}
\label{galfield}

Given the fact that large-scale magnetic fields in vacuum would be direct
evidence for a limit on their exponential decay with distance (and hence a limit on the photon mass), large-scale magnetic fields in the galaxy or even in extra-galactic space have long been of interest.   Yamaguchi \cite{yamaguchi} wrote the pioneering comment, arguing that turbulent cells in the Crab nebula of size 0.1 ly = $10^{15}$ m implied a Compton wave length of at least this size:
\be
\lc   \gtrsim 10^{15}~ \mathrm{m} ~~~ (0.1 ~\mathrm{ly}),
\ee

One can start to evaluate this claim by looking at magnetic fields through measurements of frequency-dependent rotation in the plane of polarization of electromagnetic waves (Faraday rotation).  The polarization rotation is sensitive to the product of plasma density and magnetic field strength, and in many cases the observations are consistent with uniform plasma and field distributions.

However, these observations also would be consistent with a volume-averaged value for the product, even if each individual factor  varied substantially.  For example, the density $\rho$, which is non-negative-definite, must have a
nonzero average, but $\mathbf{B}$ might have zero average, even with $\langle \rho \mathbf{B}\rangle$ nonzero.  Thus, as a matter of logic, the nonzero average of $\langle \rho \mathbf{B}\rangle$ does not have any unavoidable implications for the magnitude of $\mathbf{A}$.

Besides Faraday rotation, an even more conspicuous signal of interstellar
magnetic fields is synchrotron radiation.  Because this radiation would look exactly the same if the direction of a magnetic field ${\mathbf{B}}$ were reversed, data on this phenomenon cannot discriminate against frequent reversals of the field, and thus are consistent with the zero average field suggested in the above
paragraph \cite{rainer}.\footnote{
A similar comment about insensitivity to field reversals applies to signals from Zeeman splitting of OH and other molecules, as well as linear polarization of interstellar dust grains.
}

The same kinds of question apply even more to limits based on galactic-sized fields \cite{wp,byrne,chib}, because observations on such scales are even less precise.  Indeed, Chibisov \cite{chib} claimed a limit
\be
\lc  \gtrsim  10^{20} \,\mathrm{m}  ~~~ (10^4 ~\mathrm{ly}),
\ee
by following Yamaguchi \cite{yamaguchi} and extending the Crab Nebula analysis to the galaxy.

If the region of uniform $\mathbf{B}$ extends over a galactic arm, and is
aligned parallel to the  axis of the arm, then $\mu A\sim \mu R B$ (where $R$ is the radius of the arm) arguably should be no bigger than $B$:
This would follow from the virial assumption that plasma kinetic energy,
ordinary magnetic field energy, and photon-mass-induced vector potential energy all should be in electromechanical equilibrium. Thus,
the energy density associated with the magnetic vector potential should not vastly exceed the energy associated with the magnetic field.

The virial assumption recently has been asserted forcefully by ADG
\cite{adel}.  They argue that if
the galactic magnetic field is in their `gauge-symmetry-breaking' Proca regime, then the very existence of a large-scale field would mean that the Yamaguchi-Chibisov limit is valid, though they propose a slightly smaller number,
\bea
\lc &\gtrsim& 3 \times 10^{19} \,\mathrm{m}  ~~~ (1\  \mathrm{kpc}), \ ~~~ \mathrm{or}  \nonumber \\
\mu &\lesssim& 6 \times 10^{-27}\,\mathrm{eV}
\equiv  10^{-62}\,\mathrm{kg}  \ .\label{chib-adg}
\eea
If one could confirm sufficiently detailed information about the plasma
and the magnetic field, such a result might become well established.
However, precise galactic field assertions are not provable today, although they might be established in the future.

Further, at present,  there are at least two obstacles, besides those mentioned already.
First, there could be significant time dependence of the fields on a scale as small as 1000 years.  Secondly,  there is good reason to believe that there are substantial inhomogeneities in the field and plasma, which could be reservoirs of much greater total energy than the average magnetic field energy.  Ryutov \cite{ddrVT} recently has examined such ambiguities, and emphasized a tacit (but not obviously valid) assumption needed for the virial theorem, that one is dealing with a closed system.

Still, the Proca energy emphasized by ADG is so large that it would be tempting to dismiss all the above caveats, and at most use them to weaken somewhat the Yamaguchi-Chibisov limit associated with phenomena on a given scale.    However, there is another issue already hinted at above which can change the calculus completely.   If the photon mass were zero, then data consistent with uniform magnetic fields over large regions naturally should be interpreted as indicating that uniformity really is present.   After all, there is no obvious mechanism for reversals, and no natural length scale for the reversals.   The same kind of energy consideration championed by ADG changes this analysis if the mass is nonzero.

With a given photon Compton wavelength $\lc$, balance of energy among plasma, magnetic field, and photon mass contributions could occur if there were "pencils" or filaments of plasma with an average $\mathbf{B}$ aligned in one direction parallel to the filament axis, and outside each filament an exponentially decaying vector potential producing an equal and opposite flux to that contained in the filament.  As explained above, such a configuration would be consistent with all observations to date relating to $\mathbf{B}$.\footnote{
Attentive readers will note that the above description in its simplest form involves discontinuities in plasma density or its derivative.  The implications would not change, while the picture would become much more plausible, if one instead assumed only continuous changes in plasma density.  Thus a region with average positive magnetic field would have a relatively large average plasma density, while the surrounding regions with average negative magnetic field would have a much smaller average density, even though nonzero.
}$^,$\footnote{
Interestingly, on extragalactic scales, there is evidence from simulations (simulations that utilize the  'cold dark matter' hypothesis) that primary fluctuations interacting with gravity might produce filaments of plasma, which naturally would be associated with magnetic fields.  Of course, in the absence of photon mass, there would be no sustained magnetic field outside such a filament.  For a recent discussion of these issues, see \cite{Refregier:2002uxa}, and for possibilities of magnetic fields in filaments see \cite{Keshet:2004dr}.
}

There is another relevant set of observations within our galaxy, the  velocity dispersion of pulsar radio signals mentioned in Sec. \ref{Kroll}.  It is proportional to the integrated plasma density along the  path between each pulsar and the observation point \cite{feinberg}.  This clearly gives a constraint on the average plasma density, but given the relative paucity of pulsars may not provide enough information to determine whether there is or is not a filamentary structure on a particular scale.

We believe that  something like the Yamaguchi-Chibisov limit might be verified in the not-too-distant future by additional observations
(thanks to extraordinarily rapid progress in gathering astrophysical data
beginning in the last decade or so).  However, it is not established by present knowledge.  There are several issues, including (besides those mentioned already) the  poorly known
magnetic fluctuations at short distance scales (tens of pc),\footnote{
These fluctuations are, however, certainly substantial compared to the uniform or slowly-varying field.
}
the role in the virial theorem of gravitational energies, and short-time phenomena that `dump' energy into the medium, especially supernova explosions.\footnote{   At least some of the relevant factors are discussed by Beck \cite{rainer}.}

When we come to galactic-cluster-sized magnetic fields, the same
problems are even more challenging, because the detail available at greater distances of course is reduced.


\subsubsection{The Lakes method}

With perhaps the most creative observational method put forth in half a century for detecting photon mass, Lakes \cite{lakes} proposed to measure the torque on a magnetic flux loop as it rotates with the Earth' surface.   Lakes noted that if a magnetic field $\mathbf{B}$ is nearly uniform over a region of dimension $L$, then at a typical random point the vector potential is of order $LB$ in magnitude.  The $-\mu^2\mathbf{A}^2/2$ term in the Lagrangian then leads to a toroidal moment interaction between a toroidal solenoid of moment $\mathbf{a}$ and the ambient ``vector potential field"  $\mu^2\mathbf{A}_{\rm amb}$,  analogous to the torque on a loop of electric current from an ambient magnetic field.  In other words, nonzero photon mass makes the vector potential observable, and this technique allows its direct observation.

The torque
\be
{\boldsymbol{\tau}} =
{\boldsymbol{\nu\times}}
{\mu^2}{\mathbf{A}}_{\rm amb}
\ee
acts on {\boldmath${\nu}$}, the `vector-potential dipole moment' of the
flux loop.  As one knows {\boldmath${\nu}$},  measuring or
limiting the value of the torque on the solenoid,  {\boldmath${\tau}$}, yields  $(\mu^2\mathbf{A}_{\rm amb})$.  Determining a lower bound on
$\mathbf{A}_{\rm amb}$ then places a value on $\mu$.   A typical value of ${\bf A_{\rm amb}}$ might be very large in galactic and intergalactic
space, when $|\mathbf{A}|\approx |\mathbf{B}|L$ with $L$  the radius of a cross section transverse to a cylinder aligned parallel to a field ${\bf B}\approx$ constant.

In his original experiment, Lakes \cite{lakes} studied the torque on the
solenoid about one particular axis (the rotation axis of the Earth), and hence had to assume that this axis was not parallel to $\mathbf{A}$.   He also assumed, based on inferred values for galactic and intergalactic fields and the associated scales $L$, a magnitude for $A$, and thus obtained a limit.

A later experiment by Luo et al. \cite{luotorsion}, both was more precise and also allowed the axis about which the torque was measured to vary in time.  This eliminated Lakes' angle problem, but still left the assumption that the magnitude of $A$ is $\langle B\rangle L$.   These experiments \cite{lakes,luotorsion} suggested that a lower limit on $\lc$ as high as $3\times 10^{11}$ m could be obtained  from fields in the Coma cluster of galaxies (Abell 1656, whose center is about 100 Mpc from Earth).  This would be even stronger than the solar wind limit.

Unfortunately, at present the assumption
$|\mathbf{A}|\approx |\mathbf{B}|L$ is not guaranteed for measurements on Earth \cite{gntorsion,luotorsion2}.  This is true not only because (as Lakes pointed out \cite{lakes}) one may in principle be near a zero of $A$, but also because the evidence for uniformity of $\mathbf{B}$ is fragmentary.  If there are holes in the distribution of the plasma supporting $B$, and if we are in such a hole, then, with a substantial $\mu$, the linearly growing $A$ envisioned by the experiments \cite{lakes,luotorsion} would be damped exponentially. Thus, $A$ could indeed be small, making even a large $\mu^2$ invisible in these experiments.

ADG \cite{adel} observed that in their vortex scenario the effective $A$ would also be much smaller than $LB$, and again only a much weaker limit would hold.  Thus, both in the Proca case and the vortex case it is not possible at this point to obtain a secure quantitative limit using the Lakes method.

There is another possible approach to seeking a value for
$\mu^2\mathbf{A}$ (as mentioned already in the discussion of Ryutov's solar wind limit):  In the presence of plasma, a static magnetic field may take exactly the form it would have in $\mu=0$ magnetohydrodynamics, {\it provided} \cite{GN,wp,gntorsion,luotorsion2}  the plasma supports a current $\mathbf{J}$ that exactly cancels the  `pseudo-current'
$-\mu^2\mathbf{A}/\mu_0$ induced by the photon mass.  Thus, a uniform average  $\mathbf{B}$ over a region large compared to $\lc$ would require such a plasma current. This holds even if there are large fluctuating fields in addition to the average field.

By putting an upper limit on the true current one would put an upper limit on $\mu^2\mathbf{A}$.  This limit would not be subject to the caveat that
$\mu^2\mathbf{A}$ may be small at some particular point, because the plasma covers the same volume as the apparent volume over which  $\mathbf{B}$ is spread. If like Lakes and Luo et al. one considers the Coma cluster, one may obtain a more conservative upper estimate of a possible plasma current, as follows.

From \cite{carilli}-\cite{clarket} there are estimates
$L\lesssim 1.5 \times 10^{22}$ m, $B\gtrsim 10^{-10}$ T, the plasma free electron density in interstellar space satisfies $\rho\lesssim 10^4$/m$^3$ and the plasma temperature $T\lesssim 10 $ keV.  Taking the generous view that the electron velocity in a coherent current could be as big as the r.m.s. thermal velocity yields a limit
$\mu^2 \langle A\rangle \lesssim 10^{-13}$ T/m,
two orders of magnitude smaller than the laboratory experimental result.
Furthermore it is unaffected by uncertainties about zeros in $A$ at any particular location (such as the Earth).  Thus observations of
volume-averaged properties of the cluster {\it could} yield a  more conservative upper bound
\bea
\lc &\gtrsim&  3\times 10^{12} ~ {\mathrm m~~~~ or} \nonumber \\
\mu &\lesssim&  7\times 10^{-20} ~{\mathrm eV} \ \equiv  10^{-55}~              ~\mathrm{kg} \  .
\eea

This ($\lc \gtrsim 20$ AU)  {\it would} be substantially better than the Lakes-method limits and the solar wind limit.   We use the conditional forms {\it could} and {\it would} because, as discussed in \ref{galfield}, there still are issues associated with the inference of large-scale uniform magnetic fields from observations.

A phenomenon that could be used as an even more conservative way to obtain a limit comes from the fact that the circulating current in the presence of ${\mathbf {B}}$ leads to a large Lorentz force, tending to `explode' the plasma.   (This was emphasized by ADG and used in Ryutov's solar wind analysis,)  A careful calculation of the rate of expansion could provide an estimate of the current, and hence yield a limit on $\mu^2{\mathbf {A}}$.  This could imply
\bea
\lc &\gtrsim& 3\times 10^{9} ~ {\mathrm m~~~~ or} \nonumber \\
\mu &\lesssim&  7\times 10^{-17} ~{\mathrm eV}
      \equiv 10^{-52}~\mathrm{kg} \ .
\eea
Albeit with all the same reservations mentioned before, this is a reduced Compton wave length of about 4 times the radius of the Sun, $R_\odot$ \cite{gntorsion}.

Recently Ryutov \cite{ryutov2009} showed that such an approach applied to satellite observations of the solar wind produces a limit on the product $\mu^2\mathbf{A}$, nine orders of magnitude less than obtained from the best laboratory experiment measuring torque on a toroidal magnet, by Luo et al. \cite{luotorsion}. 


\subsubsection{Pursuing the Higgs effect}

Among other points they discussed,
ADG \cite{adel} noted that if $\mu$ arises from an Abelian Higgs mechanism in the regime analogous to that of a Type II superconductor, the existence of a  non-zero photon mass  implies generation of a primordial magnetic field in the early universe.   This is quite interesting, because in the absence of any photon-mass considerations there has been substantial debate in the astrophysical community about whether the galactic field had a primordial `seed' or is solely a consequence of a currently existing `galactic dynamo'.

There remain  significant issues for the Higgs scenario.  Indeed, at present no explicit  or feasible  
value
has been proposed for a Higgs photon mass limit. A theoretical basis for the physical parameters ($q.\lambda,v$) needed to make
the vortex idea workable is lacking.  Clearly the parameters would be enormously smaller than for the still unverified electroweak Higgs.
In view of the many very small ratios of parameters found in
particle physics already, this is not absurd, but it also is not compelling.

As discussed  in Sec. \ref{galfield}, given the complexities in the real astrophysical world, it may not be easy to distinguish effects of those complexities  from effects of Higgs vortices.  The flood of  new data which we can confidently expect in the relatively near future may well shed light on these issues by further clarifying the properties of astrophysical structures.


\subsection{Photon dispersion as a lead into gravity}

We explained earlier that limits on photon mass from static fields already
are so stringent that any consequent observable dispersion in photon velocity  likely is ruled out.\footnote{
Although as we have seen in \ref{Kroll} not by as enormous a factor as holds for dispersion of pulsar wave velocities which we discuss now.
}
Nevertheless, there {\it is} an observed dispersion with frequency in arrival times of electromagnetic waves from a pulsar to a detector in our vicinity.  If this is not due to a photon mass, one has to determine another cause, and the obvious one is interaction with the interstellar plasma \cite{feinberg}.  In fact, this ``whistler effect" gives a way of detecting the mean plasma density along the path of the pulsar signal.

The phenomenon introduces a notion that will become even more important in the discussion of gravity to follow:  When deviations are found from the
implications of theory with known sources taken into account, one must look for modifications in the theory, or additional sources (or, of course, both).  In the radio dispersion case the plasma explanation fits so many facts so well  that there is no controversy about it, no suggestion that there is something missing in Maxwell theory.

This ``non-mass" source of photon velocity dispersion has special interest for us,  It was the effect  \cite{feinberg} that first enticed us to study the photon-mass issue \cite{mmncline}, at the time of the early pulsar discoveries.  In the discussion of gravity to follow, the question of whether to ascribe anomalies to modification of  gravity or to the addition of sources will become  more interesting.


\subsection{The primacy of length over all other measures}

After people had considered the old, esthetically motivated 
scaling notion of a power deviating from that of the inverse square law (introducing no specific length parameter), 
they came to a physically motivated idea, giving a nonzero mass to the photon.  Very early it became clear that the only likely observable effect along these lines would be a departure from Maxwellian structure for the very long-range behavior of static fields.  By now the length scale in question is related to solar system dimensions, and there is every reason to expect that it will be extended much further still.

In principle, the Abelian Higgs formulation might accommodate (for true vacuum at zero temperature and zero ambient magnetic
field) an actual, observable photon mass giving measurable dispersion of photon velocity  with energy, but that is (literally) quite remote from anything we might hope to detect.  ADG suggested that perhaps beyond galactic scales, where magnetic fields are somewhat weaker than in our galaxy, a finite, even directly observable photon mass might emerge.   It could be interesting, and certainly would be challenging, to find  types of observation that could be sensitive to such an effect.


\section{Gravitational theories}
\label{grav}

There are interesting parallels as well as divergences between the developments of electromagnetic and  gravitational  theories from their initial formulations with static forces acting at a distance to the eventual construction of dynamical fields.
The most generic statement is that the latter has evolved more slowly.  It began earlier, but even today it is less developed and also (in some important ways) less well-tested.    Gravity as a theory began `instantly' with Newton's $1/r^2$ force law.  However, after that, despite burgeoning successful applications, it remained literally static until Einstein's general relativity more than two centuries later, with its dynamical theory for the gravitational field.

The realization that there are wave solutions of the equations of electromagnetism arose   in the mid-nineteenth century (less than a century after Coulomb's law), but the analogous statement for gravity came only with the advent of GR.   Even after that there was wavering for at least half a century about the existence of these waves.   By that time, quanta of electromagnetism, photons, were long established, so that wave and particle properties of light were on an equal footing.

Also, perturbative quantum electrodynamics [QED] had become a science still being refined today.   The quanta or particles corresponding to gravitational waves are unlikely to be observed in the foreseeable future, simply because of the extraordinary weakness of gravity at scales accessible to humans.  Indeed, even the existence of classical gravitational waves has been established only in the same sense as for neutrinos in the first half of the 20th century:  Their radiation accounts quantitatively for energy loss observed in binary pulsar systems.  Absorption of energy from gravitational waves, yet to be confirmed,
is a target of  current and planned large-scale gravitational-wave detectors.

Meanwhile, a quantum theory of gravity analogous to QED does not exist,
in part because the most straightforward formulation is not renormalizable.   String theory offers promise of providing a consistent quantum formulation including gravity, but still is far from complete.

It should not be surprising that, even more than in the case of
electrodynamics, long-distance, low-frequency deviations from the preferred theory are more likely to be detected in the study of quasi-static phenomena than in an effect like frequency or energy dispersion of wave or graviton-particle velocity.  Once again, let us review the stages in evolution of the subject.


\subsection{Newton's law of gravity}

According to Newton, the force between two masses acts along the line between them and takes the form
\begin{equation}
F=-\frac{Gm_1m_2}{r^2}  \ \ .
\end{equation}
The success of this form was the basis for the later introduction of Coulomb's law.  Here the negative sign indicates that the force between two masses is attractive, unlike the repulsive force between like-sign electric charges.

Of course, even at an early stage celestial mechanics gave a much higher
precision in verifying the inverse-square law for gravity than the corresponding law for electricity.\footnote{
The first quantitative test for the inverse-square law of electric force was done by John Robison in 1769, predating Coulomb  \cite{sciam}!  It yielded an accuracy of $F \propto r^{-(2 +q)}$, where q was found to be $\sim 0.6$ on a scale of a few inches. (Robison ascribed the $0.6$ to experimental error, but the precise use of numerical uncertainties awaited the seminal inspiration of Gauss' work on least squares in 1801.)    Contrariwise, at the end of the 1500's Tycho Brahe's naked eye observations were already good to 1 arc-sec or better \cite{astronency}, about the naked eye diffraction limit of $\sim\lambda_\nu/D \sim 10^{-4}$.  Kepler used these observations to establish his laws of planetary motion, specifically the ellipse for the Mars orbit.  Half a century later Newton quantified this in the inverse-square law, with the advent of telescopes bringing increasing accuracy \cite{astronency}.
}
Newton himself considered $GM$, which is much easier to measure than $G$,
what we now call Newton's constant.\footnote{
Even today, $GM_\odot$ for the Sun is known to a part in $10^{10}$  whereas $G$ only is known to about a part in $10^4$ \cite{astroq}.
}

Newton never reported an attempt to determine $G$, even though he had built pendulums of size 11 feet and had correctly calculated the average density of the Earth to be about 5-6 times the density of water \cite{g1,g2}.  The reason for his omission   appears to be a surprising error that appeared in the Principia, stating that two spheres of Earth density and of size one foot placed 1/4  inch apart would take of order a month to come together, indicating that terrestrial experiments would be useless.

As discussed by Poynting \cite{poynting}, Newton's error produced an inhibition against performing terrestrial experiments until the work of Cavendish \cite{cavendish}.   Cavendish's purpose, the same as Newton's stated goal, was to determine the average density of the Earth.   For this he needed only the ratio of the gravitational force between two test bodies of known mass to the
gravitational force exerted on a test body by the Earth.  He did not explicitly compute or even define G, which was introduced only much
later.\footnote{
An early reference to measuring ``$G"$" was given by Cornu  and Baille \cite{cornu} (who called it ``$f$").  In some folklore Cornu is given credit for popularizing the use of ``$G$".
}

Over the following century, advances in mathematics allowed ever more precise calculations, and Newtonian theory always triumphed.  Then, in 1781 Herschel discovered what he first thought was a new parabolic-orbit comet, but which quickly turned out to be a new elliptical-orbit planet, Uranus.  (The entire story is described in \cite{grosser}.)  In 1784 Fixlmillner combined two years of then modern observations with two old sightings that had been mistaken for stars and calculated an orbit.  By 1788 this elliptical orbit already did not work.

By 1820 there were 39 years of recent observations combined with 17 ancient observations (going back to 1690).   Bouvard used these data to calculate a precise orbit but could not reconcile the entire data set.  To resolve the dilemma he specifically attributed gross error to the ancient observations of eminent astronomers rather than allow for some unexplained cause of the irregularities.   This all led to much disagreement, and over the succeeding decades the observed deviations from calculated orbits got worse.

Into this situation came John Couch Adams and Urbain Jean Joseph le Verrier. In the time frame of 1843-1846 they independently used Newton's Law to predict the location of a new planet, Neptune, discovered in 1846 by Galle, on the first day he looked \cite{grosser}.  They solved what we would call an inverse problem:  What object was causing the not-understood perturbations of the planet
Uranus?\footnote{
An input into the solution \cite{grosser} was what amounted to the Titius-Bode Law of Planetary Distances \cite{n1,n2}.
}

Clearly the  Neptune  solution (what in today's parlance would be called `dark matter')\footnote{
Another such problem was announced by Bessel in 1844 when he concluded that the observed wobble in Sirius' location must be due to a companion
\cite{bessel}.  (He made a similar observation for Procyon.)  In 1862 Alvan Clark observed the very faint companion.  We now know that it could cause the wobble because it is a high-density white dwarf.
}
also had an alternative explanation, a modification of gravity.  A similar issue arose soon after when le Verrier started a complete study of all the planets.  When he returned to Mercury in 1859, he again found an earlier troubling problem \cite{leV2}, the precession of Mercury's perihelion was too large, by 33-38 arc-seconds/century
\cite{leV2,leV,vulcan}.
Later Simon Newcomb did a more precise calculation and found the ``modern" value of 43 arc-seconds/century.\footnote
{See. p. 136 of \cite{vulcan}.
}

The ``obvious" most likely resolution was that there had to be a new planet, Vulcan, in the interior of the solar system. However, this time the answer was {\it not} missing dark matter.  A hint in that direction was given by Asaph Hall.  He followed on Newcomb's observation and Bertrand's work (which led to Bertrand's Theorem).  Bertrand had
shown that for small eccentricity the angle between successive radii vectors to the closest and furthest points in a bound orbit is \cite{bertrand}
\be
\theta = \frac{\pi}{\sqrt{n +3}},
\ee
where $n$ is the power law of the force ($n=-2$ for Newton's law).

Using this, Hall \cite{hall} calculated that a force law with $2\rightarrow 2.000~000~16$ would account for Mercury's precession.\footnote{
Amusingly, this is precisely an example of the original way of parametrizing departures from the Coulomb law, thus very much in the spirit of Hall's time.
}
Of course the accepted resolution today is the replacement of Newtonian gravity by general relativity, effectively leading to an added (small) $r^{-3}$ term in the force law.

It is worth dwelling on this a bit.   Hall's parametrization was a purely
phenomenological one.   It is hard to imagine that a phenomenological approach could ever have come close to evoking the complete general relativity.  However, a discrepancy like that of the Mercury orbit was an alert for a possible need to modify the theory, and did give  guidance for a possible (though in the end incorrect)
form of the required modification.\footnote{
During the same period, another modification of gravity was proposed by Tisserand to explain Mercury's orbit \cite{whittaker,tiss1,tiss2}.
The idea was to add velocity-dependent terms, as Weber had done for electricity.   In more modern notation the force can be described by
\be
F = -\frac{Gm_1m_2}{r^2}\left(1 - \frac{\dot{r}^2}{2c^2}
              + \frac{r \ddot{r}}{c^2}\right).
\ee
However, as modern work shows \cite{bunchaft}, such a force cannot be conservative and also explain both Mercury's perihelion shift and the deflection of light by the sun.
}


\subsection{Einstein's general theory of relativity and beyond?}
\label{Egr}

While the evolution of electrodynamics entailed a harmonious progression fed both by experiment and by theory, the next stage in gravity was a theoretical accomplishment.  General relativity [GR] immediately provided an accurate  solution to the Mercury precession problem.   Soon GR was vindicated by observations of the solar deflection of light, and more recently has been vetted by many other tests.  Einstein's eight-year intellectual struggle, assisted by many colleagues, produced general relativity as a new version of Newton's gravity, now consistent with the principle of relativity, and constituting a dynamical field theory like Maxwell's electrodynamics.

Given the assumption that gravity is a metric theory, a systematic
parametrization of such theories for phenomena depending on gravitational
sources with velocities low compared to the velocity of light yields the PPN or parametrized post-Newtonian expansion for corrections to Newtonian gravity.  Einstein's minimal theory, with no added gravitational fields besides the metric itself, gives definite values for these parameters, and many observations have provided increasingly stringent limits on deviations from the Einstein values.  Because this subject has been reviewed extensively in the literature \cite{will1,will2}, we refer the reader there rather than sparingly touch on the
same material here.

Although the PPN program began as a search for a certain class of deviations from Einstein gravity, as with scalar-tensor theories, it really has become more an increasingly extensive set of verifications for GR.  As such, PPN so far has followed a similar trajectory to the search for photon mass described earlier in this paper -- much interesting and creative theoretical work, many beautiful and ingenious experiments, but no evidence of any deviation from the simple starting point.

Finally, there is the school of quantum
gravity,\footnote{
A question that also comes from the school of quantum gravity is whether there are measurable vector and scalar partners of the graviton that have mass; so-called ``fifth forces." These would die out after a finite distance, leaving only the effects of the zero-mass graviton behind.  Repeated experiments, on scales from the laboratory to astronomical, have thus far found no evidence for such forces \cite{fifth}.
}
whose most intensely studied formulation in recent decades is string  theory \cite{stringphoton}, with its predicted extra dimensions \cite{extra}.  Besides the long-distance deviations on which we focus here this can also produce deviations at short distance scales and in strong gravitational fields.  These both are not easy to detect, though the former at least may be subject to laboratory investigation.


\subsection{Nonzero graviton mass:  Is it possible?}

\subsubsection{Early considerations}

A naive approach to modifying gravity at long distances would be imitate Proca and introduce a massive graviton analogous to the massive photon.   This could be meaningful even though individual gravitons may never be observable.  It turns out, however, that the intricate structure of GR makes introduction of a graviton mass a much more delicate exercise.  The upshot is, as we shall see in the following, that a graviton mass corresponding to a length scale much smaller than the radius of the visible universe appears to be excluded.  Certainly any corresponding velocity dispersion would be unobservable.

The study of long-range deviations from GR in this context began in 1939 with  papers  by Wolfgang Pauli and Markus Fierz [PF] \cite{PF,FP}.  Fierz was Pauli's assistant, and this was his ``Habilitation" thesis.  They considered particles with finite mass, which meant that in the rest frame of such a particle with spin $s$ there must be $2s+1$ degrees of freedom.  This is in contrast to the two degrees of freedom (helicity $\pm s$) for 
neutral 
massless particles implied by CPT invariance (Of course, for helicity zero, there is only one state). 

For an integer-spin particle with spin wave function represented by a
contravariant tensor, one obtains the constraint $\partial_\alpha
T^{\alpha\beta........}=0$, meaning that in the rest frame the spin wave
function is described by a tensor with no time components.\footnote{
For a massive spin-one particle this is just the Lorentz gauge condition discussed earlier for a massive photon.
}
This tensor should be symmetric under interchange of any pair of indices, as well as  traceless in any pair.   Simple counting shows that these conditions give $2s+1$ degrees of freedom if the tensor has $s$ indices.

The focus of PF was on coupling of these massive particles with spin to the electromagnetic field, {\it not} on speculating about a massive
graviton.\footnote{Thus
they were following closely Proca's original approach in his papers on a massive spin-1 field.
}
Indeed, there are difficulties with  minimal electromagnetic coupling for  single-mass single-spin wave equations with spin higher than  $s=1$.   In particular, Rarita and Schwinger  \cite{Rarita:1941mf} looked at these issues for
$s=\frac32$.  It was not until the 1960s that such matters gained serious
attention.  This eventually led to a consistent theory for spin-one charged particles, identified first with $SU(2)$ and then with $SU(2)\times U(1)$ non-Abelian gauge theory for electroweak physics, as discussed in \ref{higgs}.  Even later, in the 70s, this kind of consideration was one of the routes that led to supergravity, and its relation to string theory.

The Pauli-Fierz approach to a massive graviton starts with the notion that
space-time is approximately flat.  Then one may consider small-amplitude
deviations and describe them by a wave equation like the Proca equation for the electromagnetic field.   The usual Einstein equation is modified by addition of a mass term:
\be
G_{\mu\nu}-m^2(h_{\mu\nu}-\eta_{\mu\nu}h) = GT_{\mu\nu}  \ \ ,  \label{pf}
\ee
where $m$ is the graviton mass in inverse-length units, $\eta_{\mu\nu}$ is the Lorentz metric, $h_{\mu\nu}=g_{\mu\nu}-\eta_{\mu\nu}$ is the departure of the metric from perfect flatness, $h=\eta^{\mu\nu}h_{\mu\nu}$ is the
four-dimensional trace of $h_{\mu\nu}$, $G$ is Newton's constant, and
$G_{\mu\nu}$ is the Einstein tensor to linear order in $h_{\mu\nu}$:
\bea
G_{\mu\nu}&=&\Box ( h_{\mu\nu}-\eta_{\mu\nu}h) -\partial^\alpha\partial_\mu h_{\alpha\mu}    \nonumber \\
&~& - \partial^\alpha\partial_\nu h_{\alpha\nu}+\eta_{\mu\nu}\partial^\alpha\partial^\beta
h_{\alpha\beta}+\partial_\mu\partial_\nu h  \  .
\label{ein}
\eea

Ignoring the details to be discussed below, we may see easily why this expression suggests a massive graviton.  Focusing on the  mass term on the right of Eq. (\ref{pf}) and the first term in Eq. (\ref{ein}) for  $G_{\mu\nu}$, we get, in explicit space and time notation, 
\be
(\nabla^2-\frac{1}{c^2}\partial_t^2-m^2)(h_{\mu\nu}-\eta_{\mu\nu}h)=0  \ \  . 
\ee
This is precisely the wave equation for motion corresponding to nonzero rest mass $m$ expressed in inverse-length units.  This equation also makes clear why one should get exponential behavior for a solution at zero frequency.

Once again, the five degrees of freedom for $h_{\mu\nu}$ with non-zero $m$ must all be present for the limit $m=0$, but this time the narrow escape for the photon discussed in \ref{phlimit} does not quite work.  The helicity-$\pm 1$ states appear with a four-gradient factor, and integrating by parts in the coupling to $T_{\mu\nu}$ yields a four-divergence which vanishes because of local conservation of energy
and momentum.  However, the helicity-$0$ state multiplies the trace of
$T_{\mu\nu}$, and this in general does not vanish.  

Thus even in the $m=0$ limit we have a scalar-tensor theory of Fierz-Jordan-Brans-Dicke type
\cite{fierz,jordan,brans}.  Ths means that gravitational coupling between masses occurs not only through the (tensor) Einstein gravitational field but also through an additional Lorentz-invariant, or scalar, field.  

Besides these points, the reader may be curious why Eq. (\ref{pf}) takes the precise form that it does.   The answer is that for any other linear combination of the two pieces the `potential energy' density is unbounded below, corresponding to a `tachyon', an unphysical  particle with negative squared mass, which if it did exist would have the paradoxical property of moving always faster than the speed of light.  

There is a general approach to dealing with tachyons, namely, looking for an appropriate background configuration about which only positive-energy fluctuations can occur.   In this case, that still would not be  satisfactory, because it would mean flat space would be excluded as a possible spacetime background geometry for relativity.   Nevertheless, there have been studies in this direction, indicating that in such a case the famous event horizon associated with a black hole might not occur.  That is, there could be mass centers which from the outside would look very much like black holes, but whose inner structure would be quite different, and less singular \cite{visser,grish}.

Thus, if one reflects on the philosophical perspective in which one might hope to contemplate graviton mass, there is the quandary that it seems necessary to pick a particular background spacetime metric against which the mass effects would be defined.   This seems to contradict at least the spirit of general relativity, which in principle could accommodate almost arbitrary spacetime geometries.  Of course, nothing like this problem occurs in the case of photon mass, where the natural background is Minkowski space.


\subsubsection{Recent considerations \label{vdv}}

Surprisingly,  the realization that there are serious problems with the zero-mass limit  took a long time coming, and then was announced in 
three almost simultaneous papers, by Iwasaki \cite{Iwasaki:1971uz} by van Dam and Veltman \cite{v2} and by Zakharov \cite{v1}.  This assertion is popularly known as vDV-Z, because the latter two papers received more early attention.]
van Dam and Veltman found the most striking aspect of their result in a
comparison with non-Abelian gauge theories with fixed vector particle mass (i.e., mass but no Higgs mechanism).   Such a
theory, as we observed earlier, is not perturbatively renormalizable. and so clearly has a discontinuity at zero mass (where it {\it is} renormalizable). The paradoxical discontinuity in a mass or inverse-length parameter found for gravity by vDV-Z occurs already in purely classical field theory.  This was the stimulus for the ensuing theoretical study of graviton mass, and also the beginning of a debate continuing till now over the viability of graviton mass as a meaningful concept.

The next stage in that contest was a paper by Vainshtein \cite{v3}, who argued that the vDV-Z position, though clearly correct in linear gravity, could be overcome by the intrinsic nonlinearity of Einstein gravity.   He started with an argument that, in the vicinity of a gravitational source, the corrections due to graviton mass should be suppressed by a factor of order $\mu^2L^2$, where $L$ is the dimension of the region around the source that is under examination.  This argument is quite appealing to  the present authors,  because we had used exactly the same notion for photon-mass effects, where the continuity of electrodynamics at zero photon mass made our argument correct \cite{GN}.  However, in {\it linearized}  massive gravity the vDV-Z discontinuity would imply that suppression by $\mu^2L^2$ does not apply.

Very soon there was a riposte to Vainshtein by Boulware and Deser [BD] \cite{v4,vN}, who gave a number of reasons to question his conclusion.  They noted that graviton mass treated as a fixed constant seems to violate general-coordinate invariance, just as a photon mass seems to violate gauge invariance.   As we have seen, there is a way around that through the Stueckelberg construction, and therefore this is not a compelling point.   The analogue of the Stueckelberg-Higgs approach for gravity was introduced by Siegel \cite{Siegel:1993sk}, and discussed more recently by  
Arkani-Hamed, Georgi, and Schwartz \cite{ArkaniHamed:2002sp}, by 't Hooft \cite{'tHooft:2007bf}, and in a different way by Rubakov \cite{Rubakov:2004eb}.  Rubakov produced a formulation in which Lorentz invariance is violated, and the vDV-Z discontinuity of the linearized theory disappears.

BD also said 
that at best Vainshtein's case was not proved, because his assumptions about behavior near the source might imply exponential growth at large radius, rather than the required exponential decay.  What makes this seem a potential obstacle to the Vainshtein construction is precisely the strong coupling to scalar gravitons, even in the zero-mass limit.  Vainshtein's hope was that nonlinearity of gravity could heal this problem.   One may express this differently as hoping one can continue in from infinity the allowed exponentially decaying behavior.   
Of course, in a purely linear theory this would lead to anomalous behavior near the source.  
In fact, BD made an even stronger statement, that inevitably there must be a sixth degree of freedom, a ghost, leading to instability of the massive theory, so that the massless limit is not just discontinuous, it does not exist.  This is closely connected to their argument for exponential growth of the gravitational field with radius.  

There the matter rested for about a quarter-century, when a new context of
higher-dimensional theories inspired by string theory  led to a concrete example with something like graviton mass, the Dvali-Gabadadze-Porrati [DGP] model \cite{Dvali:2000hr}.  In this framework, our four (i.e., three plus one)-dimensional world is embedded in a five-dimensional spacetime, with the gravitational action having two pieces, one confined to our world, and the other uniformly defined over the entire five dimensions.   The fifth (purely spacelike) dimension is perpendicular to what then is a `brane' describing the three spatial dimensions of our world.

Interestingly, a group including Vainshtein \cite{Deffayet:2001uk} made the first study using this model for the gravitational field of a massive source, thus giving some vindication for Vainshtein's original position.\footnote{However,
it is not clear that such vindication is possible for the fixed (PF) mass discussed in the early work.  Once again, it appears (as in the Higgs mechanism for gauge theories) that additional degrees of freedom may be needed to provide a consistent realization of mass.
}
For a gravitational source sufficiently dilute that one may work to first-order in the mass density of the source, Gruzinov \cite{gruz} obtained a perturbative solution for the gravitational field.  This  later was made exact (though still first-order in the source) by Gabadadze and Iglesias \cite{Gabadadze:2004iy}.  The solution involves, instead of exponential decay with radius at spatial infinity, a power-law falloff including the scalar component mentioned earlier, but at sufficiently short distances it looks like the Newton-Einstein field.

Dvali \cite{Dvali:2006su} recently has explained from a very general perspective how nonlinear coupling can make the zero-mass limit continuous at finite radius. The crucial point is that the {\it source} be treatable as linear, despite the nonlinear nature of Einstein gravity.  This nonlinearity, well-known in the zero-mass case, persists when gravity is modified to include something like a mass $\mu_g\equiv r_c^{-1}$   (or some other modification setting in at or above the length scale $r_c$). 

This means that in the neighborhood of the source the field is linear in the source strength, while the nonlinearity of gravity itself suppresses the contributions of the three extra polarizations (in particular the `helicity-zero' contribution) expected for finite mass.\footnote{  It is possible that for a strong source, such as a black hole, the conclusions would be different, but perhaps only at very long times after formation of the black hole.} 
It's important to note that these discussions all employ classical field theory, so that possible quantum fluctuations, associated with the ghost field arising in the BD discussion of mentioned earlier,  could undermine the conclusions.  In this connection, Dvali observes  that for a Minkowski background metric the DGP system is ghost-free.


\subsection{More perspectives on gauge and general-coordinate invariance}

As discussed earlier in this article, gauge transformations and general-coordinate transformations are intrinsically different from the transformations associated with conventional symmetries.  The difference is that {\it no} observable is changed by such a transformation.   In contrast, e.g., rotation of an object clearly changes its orientation, so that only if the object were spherically symmetric down to the finest detail would one be unable to detect that such a rotation had taken place. 

In the following we discuss both gauge invariance and general-coordinate invariance together, because, as we shall see, they have much in common. 
A gauge transformation can be described as a rotation (or generalized rotation) in an abstract space, dependent in an arbitrary way on position in space-time
For example, in electrodynamics the transformation is simply multiplication of a complex wave function by a position-dependent phase factor. (Multiplication of a complex quantity by a phase factor is equivalent to rotation of a vector in a real, two-dimensional space.)
This is a generalization of the well-known invariance of a set of quantum states when the corresponding wave functions all are multiplied by the same (constant) phase factor.  

If {\it all} objects in some physical system were of the type mentioned in the first paragraph, i.e., completely rotationally invariant, then the dynamics of the system would be unaffected by arbitrary (different) rotations of each object's orientation coordinates.  Thus such a fictitious system gives a `concrete' model to help visualize the  meaning of gauge invariance, which would correspond to a  limit with a continuously infinite number of spherically symmetric objects. 
A gauge transformation then might seem quite different from a general-coordinate transformation, because for that the coordinates being transformed correspond to points in a physical manifold,  four-dimensional space-time itself.  

However, on reflection one might be persuaded that this perception of difference is just a prejudice.   Suppose two people trying to study motions on the surface of the earth were to use different projective maps of this (roughly) spherical manifold onto a plane, say, the Mercator projection  for one, and an equal-area conic projection using gores for the other.  Both people could describe, and even predict, the same motions, but no third party could divine, without being told, which projection each had used. 

Thus, a choice of coordinates is a mapping to a space topologically equivalent to the ``actual" or ``physical" manifold under consideration.  However, the space described by the map is unobservable, in the sense that its precise form makes no difference to observations or predictions for motions on the surface of the earth:\footnote{
This is true even though the map itself may be a physical system, such as a rendering of the Mercator projection on paper.
}  
Though particular physical points map into different points in different maps, any one of those maps may be used to describe a sequence of physical points, and will yield exactly the same sequence as any other map.

If one accepts this argument, then general-coordinate invariance involves transformations on spaces every bit as abstract and ``unreal" as those associated with gauge invariance.  Therefore, for both types it becomes intuitively obvious, or at the very least an immensely plausible conjecture, that the invariance can never be broken, not because it is a symmetry of nature but because it is an arbitrariness  of the choice of description of nature.  The modern term ``reparametrization invariance" brings this home quite clearly.  

This very simple reasoning is possible only with the benefit of decades of hindsight.   For example, Einstein was attracted to the idea of ``general covariance"\footnote{
The term refers not only to  the invariance of observations under coordinate transformations, but also to further quantities which are nontrivially transformed (and therefore not directly observable) such as vectors and tensors, as well as spinors.
} 
in 1912, but became doubtful in 1913, only returning to it in his triumphal push to the final form of his theory in 1915 \cite{stachel}.  By the end of the process, the idea {\it had} become ``obvious," but the struggle to get there made its subtlety equally obvious.  In the case of gauge invariance, as late as our 1971 review  we ignored Stueckelberg's construction.  We simply quoted a still commonly found statement that inclusion of a photon mass implies 
gauge invariance is broken and the Lorentz gauge is imposed.  Stueckelberg showed that this is false:   The Lorentz gauge is especially simple because in it the Stueckelberg field has constant phase, but there is no requirement to make this choice.  

Besides the examples we discussed earlier in this article which maintain gauge invariance even if naive consequences such as zero photon or graviton mass do not hold, one might imagine many other possible modifications.  The general argument given above implies that no matter what modifications we try, gauge invariance or general-coordinate invariance will not be broken.  Wigner wrote a famous article about the seemingly miraculous effectiveness of mathematics in science \cite{Wigner}.     In the present  context, the miracle, it seems to us, is that one can find a coordinate set to describe a physical system.   Once that is possible at all, the notion that one could choose a myriad other possible sets (with no change in physical consequences) seems easy to accept.   

Of course, different choices of coordinates (or gauges) can be useful and simple for different purposes.  Making {\it some} choice is at least convenient, and may be necessary to calculate results, as often is true for gauge choices in perturbative quantum field theory.  One should make a distinction here between coordinate invariance (or more precisely ``coordinate-choice independence") and the ``coordinate-free" formulation of a theory.  The former clearly is necessary, but may or may not be sufficient for the latter to be possible.  When it is possible, one is able to focus directly only on observable quantities, but there may be a price in terms of losing details that help to give insight into  structure.  

If one accepted that a coordinate choice is every bit as abstract as a gauge choice, a natural thought might be that gauge choices could correspond to maps of a space associated with extra dimensions, beyond those of our perceived four-dimensional space-time.   Exactly such an idea was introduced for the description of electrodynamics involving a fifth dimension by Nordstr\"om  \cite{nordstrom} even before Einstein completed his theory of general relativity.  Nordstr\"om's own, nonmetric, theory did not survive, which may help to explain why his introduction of a fifth dimension often is ignored.   

Indeed, before modern quantum mechanics, Kaluza \cite{kaluza} proposed the same idea to describe 
electrodynamics in the context of general relativity.   The works we have mentioned earlier by Klein \cite{OK} and Fock \cite{fock} came right after the development of quantum mechanics.  They were perhaps the first that may have been motivated by the  considerations given at the beginning of the preceding paragraph.   Today the role of possible extra dimensions is a hallmark in the study of string theory.

Besides pointing towards a way to unify fundamental physical theories, considerations of gauge and general-coordinate invariance also have become part of a strengthening interface between mathematics, in particular geometry and topology, and physics.   An early development was an influential ``dictionary" produced by Wu and Yang \cite{Wu:1975es}, relating gauge field theory to the theory of fiber bundles.  This gave a new way for physicists to view what they were doing, and also led to insights in mathematics generating new conjectures and proofs.  The fiber-bundle approach is geared towards coordinate-free characterization of a space, giving it an appealing generality but also making it less obviously useful for direct application in perturbative quantum field theory. For non-perturbative issues, however, this approach can be quite powerful.

We see that the word ``perspective" has a second meaning in the context of this discussion.   Part of the way we deal with the world is to process information in the form of images or maps generated from ``raw" data.  For example, photons impinging on the retina lead to perceived images in the mind.\footnote{ 
Also, interactions with other people help form one's perspectives on their psychology, as well as on social structure in general.
}  
Thus, we always are thinking and acting on the basis of our perspectives.

A gauge or coordinate choice is a perspective with which to view not only geometry but also dynamics.   These choices might be described as mathematical counterparts not only to the images we process all the time, but also to perspectives employed by visual artists through the ages.  Perturbative gauge theory is at least a craft if not an art, and its practitioners tend to have favorite choices of gauge, such as ``Feynman gauge" or ``light-front gauge."  Like the visual arts, gauge and coordinate-system choices span a gamut from maximally regular and symmetrical to highly distorted forms, yielding different sorts of insight about the objects studied.\footnote{ 
This could make the subject an interesting forum for studies of relationships between art and science.
}


\section{Phenomenology of deviations from GR}
\label{gravmass}


Given the afore-mentioned problems in even defining the concept of graviton mass, one might be tempted simply to ignore the issue phenomenologically.  Fortunately, as physics remains an experimental science, physicists have continuously attempted to parametrize a concept such as graviton mass, even, if necessary, in the face of deep theoretical problems.  These results can be very illuminating, and we discuss them in this section.

\subsection{Inverse square law}
\label{inveresegrav}


\subsubsection{Inverse square law on large scales}

Some time ago it was pointed out that looking for the largest scales over which gravity is known to work is not only a test for  dark matter but also, in the paradigm of a Yukawa-like fall off, a limit on a `Proca-type' rest mass \cite{GNgraviton}.  Even in 1974, when the Hubble constant stood at $H= (55\pm 7)$ km/(s Mpc), a conservative bound for galaxy clusters of size 580 kpc (vs. already known clusters of size 10 Mpc) yielded a bound of
\be
\mu_g  \lesssim 2 \times 10^{-65}\, \mathrm{kg}
 \equiv  10^{-29}\, \mathrm{eV}.  \label{clustermass}
\ee
This corresponds to a reduced Compton wavelength of
\be
{\lc}_g  \gtrsim  2\times 10^{22} ~ \mathrm{m}
~~~~~~~{\cal O}(10^{-4} ~ R) \ ,
\ee
where $R=c/H$ is the Hubble radius of the universe.  Slightly earlier Hare \cite{hare} had discussed enormously less sensitive limits associated with massive graviton decay to two photons and dispersion of gravitational wave velocity.

This early phenomenological work existed in the 1970's-1980's milieu of the early ideas of quantum gravity.  As described in \cite{NGanti}. a number of people realized that there could be both scalar and vector partners to the graviton, and that these partners could be massive.  Early work of Joel Scherk \cite{scherk} appears to have been especially influential in this regard.

When combined with geophysical results indicating a variation of G with distance \cite{stacey} on the scale of many hundreds of m, the stimulus was there for improved tests of gravity at all scales, be they interpreted as tests of G(r) or of new components of gravity.  This set the stage for the ``fifth force" ideas \cite{fischbach}, which originally envisioned a new force proportional to hypercharge.  This idea evolved into interest in tests of any (including new) components of gravity with Yukawa length scales from the lab \cite{luther} out to planetary distances \cite{talmadge}.  .
On the scale of the solar system, deviations are limited to about a part in $10^{8}$ \cite{talmadge}.

With the caveats of a few so-far unexplained anomalies (Pioneer \cite{pioprl,pioprd}, flyby \cite{flybya,flybyb}, variation of the AU with time \cite{AUt}), there have been no unambiguous positive results.  In this vein, astronomical searches for local dark matter have been undertaken \cite{NTAdrag,adler}.


\subsubsection{Small/large extra dimensions and ``fat/thin" gravitons}

Kapner et al. \cite{smallsize} recently conducted torsion-balance experiments to test the gravitational inverse-square law at separations between 9.53 mm and 55 $\mu$m.  This probed distances smaller than the ``dark-energy length scale" of
\be
d=(\hbar c/\rho_d)^{1/4} \approx 85~ \mu\mathrm{m}.
\ee
They found with a 95\% confidence level that the inverse-square law holds down to a length scale
$\lambda=56~ \mu$m. They also determined that an extra dimension must have a size $\hat{R}\leq 44 ~\mu$.  (Also see \cite{fatfat}.)

Note that this extra dimension should not be confused with that in the DGP
model, which is infinite in extent.  The length scale in that model comes from the relative normalization between the five-dimensional and the four-dimensional contributions to the gravitational action.  What we are talking about here actually would be a modification of gravity at small distance scales.  As such it would be a departure from the main thrust of this paper, although related to the fifth force ideas of finite-sized new forces.

A very different tack, perhaps somewhat closer to DGP, is taken in the model of Kogan et al. \cite{kogan}.  Here the extra-dimensional physics leads to graviton partners.  The first has a very small mass and the others have large mass.  Hence, the others are ``fat" and can be ignored on cosmological scales.  The first partner has a very large-distance Yukawa cut-off, but even so, below the cut-off it ends up dominating ordinary gravity.

Hence, this mass is not a graviton mass as we ordinarily think of it.  The ordinary graviton still has no mass.  Even so, one can ask if there is any large distance experimental indication of an effect of a mass for this graviton-partner.  Choudhury et al. \cite{choud} looked at lensing data \cite{weaklens} to place a limit of
\be
{\lc}_{\hat{g}} \gtrsim 100 ~ \mathrm{Mpc},
\ee
where $\hat{g}$ is to show this is for a graviton partner.


\subsection{Speed of gravity}
\label{cgrav}


\subsubsection{Conceptual questions}
\label{c-concept}

In 1799-1825 Laplace published his five volume masterpiece,
{\it M\'ecanique C\'eleste}, which transformed the study of celestial mechanics from Newton's geometrical viewpoint to one based on the calculus.   One very important point he brought up concerned Newton's (instantaneous) action at a distance.      Indeed, based on lunar perturbation theory Laplace  (incorrectly) thought that (what we would call) the velocity of gravity must be at least one hundred million times that of light \cite{whittaker,laplace}.

This question remained an open one for a century, until the triumphs of first special relativity and then general relativity led to the now standard assumption that the ``limiting velocity"  of travel for disturbances  in general  relativity, $c_g$, is the same as the velocity of light, $c$.   For long this assumption was a philosophical one, nut subject to much precise experiment.   But recently this has changed, both from the standpoint of what theory to use and also because there are now proposed methods of direct experimental inquiry.

In 1980 Caves \cite{caves} pointed out that in Rosen's bimetric theory \cite{rosen1,rosen2}, where there is both a Riemannian tensor describing the true gravitational field and a flat-space metric tensor describing the inertial forces, it is possible for the speed of gravitational radiation, $c_g$, to be less than the speed of light.  This results because $c_g$ is determined from both nearby distributions of matter and cosmological boundary conditions.   Further, massive particles are limited to velocities less than $c_g$.  Thus, in 
this theory, observations of 
$10^{10}$ eV protons  limit 
$(c-c_g)/c$ to be less than $\approx 10^{-21}$. 

In a more modern context, Moore and Nelson \cite{moorenelson} considered higher-dimensional models, where the standard model particles are confined to the standard 3+1 dimensional ``brane" but gravity can also propagate in the bulk of extra dimensions.  This leads to a $c_g$ that can be less than $c$.   Given that cosmic rays have an extragalactic origin, Moore and Nelson find a limit 
on the deviation from $c$ 
of the velocity of gravitational Cherenkov radiation 
$c-c_g < 2 \times 10^{-19} c$.

It is here appropriate to mention an idea on the speed of gravity which deals with the upcoming major theme of Sec. \ref{dmvsmg},  dark matter vs. modified gravity.   Desai et al. \cite{desai} observe that theories which try to mimic the effect of dark matter with a modification of gravity can be treated as having two metrics.  In these cases small amplitude gravitational waves couple to the metric $g_{\mu\nu}$ produced by general relativity without dark matter.  Ordinary matter, however, couples to the metric $\tilde{g}_{\mu\nu}$ that is produced by general relativity with dark matter.  The end result is that if there is a supernova signal that reaches Earth after passing through dark matter, then the gravity wave will arrive measurably sooner than, say, the light or neutrino signals.   
Perhaps with current and planned
gravitational wave interferometers  
this could be tested.


\subsubsection{Dispersion in gravitational waves \label{diss}}

Recently, a small industry has arisen based on the possibility of finding
dispersion in gravitational waves.  The starting point is the observation that, at least in some linearized theories, one can allow a massive graviton which would propagate freely via the Klein-Gordon equation of a particle with mass $\mu_g$.  If the graviton had a rest mass, the decay rate  of an orbiting binary would be affected \cite{damourpulsar,taylor}.  As the decay rates of binary pulsars agree very well with GR, the errors in their agreements provide a limit on a graviton mass.

Finn and Sutton \cite{finn} applied this idea in a reinvestigation of the data from the
Hulse-Taylor binary pulsar and from the pulsar PSR B1534+12.   From their analysis of the data they found a limit
\bea
\mu_g &\lesssim & 7.6 \times 10^{-20}~ \mathrm{eV}
\equiv 1.35 \times 10^{-55}~\mathrm{kg} \nonumber \\
{\lc}_g & \gtrsim & 2.6 \times 10^{12} ~ \mathrm{m}
\eea
to 90\% confidence level.\footnote{Note
that this value is dramatically lees restrictive than  that found by looking for departures from the inverse-square law quoted in (\ref{clustermass}).
}
This corresponds to a value of $v_g$ whose deviation from  $c$ is limited to 
roughly  a part in a thousand \cite{finn}, at a frequency comparable to the orbital frequency of the binary pulsar system.\footnote{
It should be noted that in the linearized theory, the vDV-Z discontinuity
applies, meaning that for finite graviton mass there should be coupling to a scalar graviton.   If the radius of the orbit changes appreciably during each cycle, then this would give a comparable contribution to the expected ($\mu_g=0$) quadrupole radiation.  
Vainshtein's strong self-coupling for scalar gravitons might prevent this. 
It is not clear to us what, if any,  effect this strong self-coupling would have on the ``standard" graviton-mass effect considered by Finn and Sutton \cite{finn}.
}

Baskaran et al. \cite{baskaran} have considered the effect on the timing of  a pulsar signal propagating in a gravitational field.  If the phase velocity of gravitation is smaller than that of light, they find that the pulsar timing is affected.  From limits on this deviation they find a graviton mass limit of
\bea
\mu_g &\lesssim & 8.5 \times 10^{-24}~ \mathrm{eV}
\equiv 1.5 \times 10^{-59}~\mathrm{kg} \nonumber \\
{\lc}_g & \gtrsim &  2.3 \times 10^{16} ~ \mathrm{m}
\eea

Because the phase velocity for a massive gravity wave, as indicated by the authors' own formulas, would be {\it greater} than the velocity of light, we do not see how they can use data sensitive to a velocity smaller than that of light to obtain their limit.   However, their mechanism involving resonance between two waves traveling at an angle to each other would seem to work just as well no matter which wave was faster.

Many related ideas have been proposed to measure dispersion in gravitational waves using interferometers or by observing gravitational radiation from in-spiralling, orbiting (non-pulsar) binaries \cite{willgw}-\cite{jonesgw}.
These should lead to stronger limits if gravitational wave arrivals can be
detected.  It seems clear that, as in the photon case, such limits never will be as strong as those deduced from quasi-static fields.


\subsubsection{Shapiro time delay and the speed of gravity}

If there were a graviton mass, then there would be dispersion of gravitons of different energies.  Intertwined with this is the fact that we tacitly assume that the ``limiting velocity" of GR, $c_g$, is exactly the limiting velocity of light, $c$.  This assumption is not just esthetically pleasing,  it also is of fundamental importance.

As mentioned in the introduction, 
an important aspect of the robustness of
scientific theories is the interconnections among different components.   If we look at the development of general relativity, then it 
seems natural to assume 
that the only possible limiting speed for any kind of disturbance is the speed of light, $c$.  It is always worth checking even the most strongly held claims, but one must bear in mind the cost associated with violations of those claims.   In this case, the rupture resulting if the speed $c_g$ for gravity turned out to be different from $c$ would be dramatic indeed.

Even with this as background, Kopeikin boldly suggested that if the speed of gravity differed from the  speed of light then it could be measured in the Shapiro time delay of the microwave light of a quasar passing close by the foreground of Jupiter \cite{kg1}.  Kopeikin claimed that the effect would be a first order correction, caused by the retarded gravity signal due to Jupiter's velocity, $v/c_g$.

However, Will criticized this assertion \cite{kgwill}.  His first statement was that retarded-potential theory would yield an effect
only to order $(v/c_g)^2$.  Motivated by this he used the PPN expansion of GR to find that the first-order correction to the Shapiro time delay is (in the GR limit)
\bea
&\Delta  &= -\frac{2Gm_J}{c^3} \ln\left(|\mathbf{x}_{\odot J}|
-\mathbf{x}_{\odot J}
\boldsymbol{\cdot}\mathbf{k}\right) ~~~ \rightarrow
                 \label{shapiro} \\
&-&\frac{2Gm_J}{c^3}
\left[ \ln\left(|\mathbf{x}_{\odot J}| -\mathbf{x}_{\odot J}
\boldsymbol{\cdot}\mathbf{K}\right)
\left(1 - \frac{\mathbf{K}
\boldsymbol{\cdot} \mathbf{v}_J}{c}\right)\right],
\label{gspeed} \\
&\mathbf{K} &\equiv \mathbf{k}
- [\mathbf{k} \boldsymbol{\times} (\mathbf{v}_J \boldsymbol{\times}
\mathbf{k})]/c.
\eea
The difference between Eqs. (\ref{shapiro}) and (\ref{gspeed}) gives the
first-order velocity correction, where  $m_J$ is the mass of Jupiter, $c$ is the speed of light,  $\mathbf{x}_{\odot J}$ is the distance vector from the observer on Earth to Jupiter's center, and $\mathbf{k}$ is the unit vector in the direction of the incoming light.

Note that ``$c$" is to be found in  Eq. (\ref{gspeed}) in two different places.  This is where the disagreement is.  Kopeikin would have the $c$'s inside the square brackets be $c_g$'s.  Contrarily, Will calculates Eq.(\ref{gspeed}) from GR with  $``c" =c$.  These terms are thus found to be the next order GR time-delay.

Therefore, Will finds that agreement of this formula with experiment is a (not too precise \cite{kgwill}) test of GR rather than a test of $c_g$. Will finds that any $c_g \ne c$ effects would only appear in the next order ($c_g^{-2}$).  Similar conclusions were drawn by others \cite{ss,carlip}.  The consensus  \cite{damour}
agrees with this conclusion, despite the continuing disagreement of the Kopeikin school \cite{kg3,kg4}.

There is a an appealing way to motivate the consensus position.  In first approximation, the Shapiro time delay is an effect on the propagation of light in an essentially static gravitational field, so that the speed of gravitational waves should not be immediately relevant.  Furthermore, if a heavy source is moving with respect to an observer, to first order in the source velocity the only change in the field shows no effect of retardation. Simply on dimensional grounds, acceleration of the source at most would give an effect second-order in the inverse speed of gravity.

In any event, a measurement was done when the quasar J0842+ 1835 passed within 3.7$^\prime$ of Jupiter on 8 Sept. 2002.  Fomalont and Kopeikin \cite{kg2}, compared
the $\sim(51 \pm 10)~ \mu$as deflection observed with the higher-order term.  They determined a value for ``$c_g$" of
\be
c_g = (1.06 \pm 0.19)~c \ .
\ee

This result is consistent both with standard GR (where any effect of $c_g \ne c$ appears only in order $(v/c_g)^2$) and also with Kopeikin's  theory, but with $c \rightarrow c_g$ inside the large brackets of Eq. (\ref{gspeed}).   Therefore, experimentally no nonstandard  result is found under either interpretation.


\subsection{Cold dark matter (CDM) versus modified Newtonian dynamics (MOND)}
\label{dmvsmg}

For more than eight decades it has been argued \cite{jeans,jeans2,kapteyn,trimble} that stars and globular clusters in galaxies, and galaxies themselves,  move as if they are being deflected by bigger gravitational forces than the usual assumptions of the mass in the above objects would imply.  Two possible explanations arise.  1) There is more (and different) matter present as a source for gravity than what we infer from both the visible radiation and also our knowledge of ordinary matter behavior.  2) The laws governing gravity are different from what Newton and Einstein would tell us.   (Of course, a combination of both explanations might be needed.)


\subsubsection{Cold dark matter}

Today there are two widely-discussed proposals of phenomena which may be labeled new sources of gravity: ``cold dark matter," whose implied effects 
(i.e., 
the pattern of anomalous observations which need to be explained)
are very well established \cite{darkM}, and ``dark energy," which in a very short time has become strongly indicated  by a variety of different classes of observation \cite{darkE,Peebles:2004qg,Doran:2007ep}.   Of course,
either or both sets of phenomena in principle could result from modification of GR rather than from new sources.

Note that proposals of such modifications are directly motivated by observation (hence phenomenological in nature) and so far have not yielded a well-agreed-upon conceptual basis in ``new" theory.
In this context, for the second phenomenon, ``dark energy," it may be almost a matter of definition whether this is a new kind of matter or a modification of Einstein gravity.   In particular, a constant cosmological term is consistent with all current data, and such a term put on the gravity side of the Einstein equations represents modified gravity, while put on the matter side it represents a new form of matter.\footnote{ Recent discussions of the difficulties that can arise in distinguishing by observation between dark energy and modified gravity can be found in \cite{Bertschinger:2008zb} and \cite{Wei:2008vw}.
}

``Cold dark matter," (CDM)  has been and remains the most  controversial 
question. A whole array of different observations over a long period of time has established the following:   If one assumes that we know the nature of matter in the universe, i.e., the standard-model particles (at ordinary energy scales) -- nucleons and nuclei, electrons, photons, and neutrinos, and one also assumes that we know how these elements combine to determine the structure of stars and the nature of interstellar gas, plasma, and dust, then Newton-Einstein gravity does not account correctly for the way that various objects are seen to move.

As the name implies, cold dark matter does not radiate anything we can see, either directly or through its interactions with other matter.  Thus, 
this matter should consist of slowly moving, discrete classical particles, interacting at most weakly with each other and with ordinary matter, so that the (astronomically) 
observable effects come largely or entirely from its gravitational influence on ordinary matter.  Because its interactions are presumed to be so weak, such matter could be very difficult to detect in the laboratory, and indeed there are no well-confirmed reports of such detection to date.  We have no direct evidence for the existence of dark matter.  Bertone, Hooper, and Silk \cite{Bertone:2004pz} have given a recent, thorough review of the evidence for dark matter and hypotheses about its form, including possible discrepancies with observation.

A significant problem for the dark-matter hypothesis 
(which has been extremely successful in accounting for observations ranging in scale from the size of the visible universe down to that of clusters of galaxies) 
is accounting for the precise patterns seen in the motions of visible stars and globular clusters in the edge regions of 
ordinary galaxies.  
As McGaugh \cite{McGaugh:2005er} has remarked, there is not as yet an accepted idea of how  and why dark matter should be distributed to produce the observed simple behavior of galactic rotation.  Hence this phenomenon, which was an original motive for introducing dark matter, remains a roadblock to full acceptance of the idea. 


\subsubsection{Proposed modifications of GR}

Looking at the unresolved application of the dark matter  hypothesis to explaining galactic rotation curves, Milgrom introduced a proposal he called ``MOND" (Modified Newtonian Dynamics) \cite{milgromold}-\cite{milgromnew}.  Milgrom discovered that the rotational velocity vs. distance curves (velocity of stars in orbit at a given distance from the center of a spiral galaxy) could be described simply and accurately by presuming that the acceleration produced by an isolated point mass $M$ transitions from a Newtonian $1/r^2$ behavior to a slower, $1/r$ falloff:
\be
a_N = -\frac{GM}{r^2} \rightarrow -\frac{\sqrt{GMa_0}}{r}.
\ee
The transition occurs smoothly near the distance where the acceleration falls to $\sim a_0$, a constant of size $\approx 10^{-10}$ m/s$^2$. 
After the transition distance, the rotational velocity from this acceration is ``flat'' or constant.  That is,
\be
v^2 = \sqrt{GMa_0} . \label{mondv} 
\ee 
In the application of the model $M$ is only composed of the ``visible" mass in the galaxy.  Here ``visible"
means both directly radiating matter and additional matter whose presence can be inferred from the patterns of radiation observed.  Thus what really is meant here is baryonic matter, meaning matter whose primary mass component is contributed by baryons.

This simple phenomenology has had amazing success in describing a large class of these galactic-rotation curves, gives the Tully-Fisher 
relation\footnote{ 
In Eq. (\ref{TF}), $v$ is the approximately constant (i.e, independent of radius) speed of objects in roughly circular orbits around the  outer regions of a galaxy with visible (luminous) mass $M$.
}  
\be
v^2\propto \sqrt{M}   \label{TF}
\ee
between galaxy rotation  curves and intrinsic luminosity \cite{Tully:1977fu} automatically, and avoids the need to calculate the amount of ``dark matter"  for galaxies on a case-by-case basis.  MOND originally was derived as a phenomenological fit for a few examples, but it has been used to test many further
galaxies.\footnote{ 
In principle  this  fit might imply a new formulation for the theory of gravity. Thus, the status of such considerations, and their analogues for dark energy, could be the most interesting current issue for possible long-range, low-frequency deviations from Newton and Einstein gravity.
}
Therefore, any successful dark matter solution has to be consistent with this phenomenology.  The success is too great to be an accident.

MOND advocates have found difficulty in describing mass distributions on the scale of galactic clusters.  Their preferred solution is that dark matter indeed is present in significant amounts, except they argue that this dark matter is ordinary baryonic matter, such as brown dwarfs, or at least standard-model matter, such as neutrinos with the maximum mass consistent with constraints from experiment \cite{Milgrom:2008rv}.

Bekenstein and Sanders \cite{bekenstein,sanders,beksan,bek} have found a way to embed the MOND scheme in a fully relativistic version, i.e., a classical field theory.
This involves the introduction of additional dynamical fields, a scalar,  vector and tensor field, all coupled  directly to gravity.  Zlosnik, Ferreira, and Starkman \cite{Zlosnik:2006sb} have found an equivalent formulation in which Einstein gravity is coupled both to familiar matter and to a dynamical vector field with an exotic but not pathological Lagrangian.  If it were not for the exotic aspects, one could readily identify this classical field as a form of dark matter,  even though clearly not the same as classical discrete particles.\footnote{These approaches all are based on accounting for the MOND phenomenology by alteration in the propagation of the gravitational field.   Milgrom also contemplated another interpretation, in which inertia would be modified for the regime of very small accelerations.  Ignatiev \cite{Ignatiev:2006rh, Ignatiev:2008qi} has proposed testing this alternative interpretation in the laboratory.
} 

In our view this formulation, which manifestly is not the sole, unique embedding of MOND into general relativity, should be taken as a proof by example that there can be classical-field dark matter which either exactly duplicates or accurately approximates the MOND phenomenology.  What that should be, and how it can be matched on to the standard particle picture of cold dark matter, become challenging questions for more investigation

We see classical discrete-particle matter giving a good account of extra contributions to gravity from the largest scale of the visible universe all the way down to clusters of galaxies.   On the other hand, classical continuous fields acting as sources of gravity neatly describe extra contributions to gravity at galactic scales and below.   In terms of their established domains of applicability, there is a conspicuous duality or complementarity between the two approaches.

The obvious question is, "How can the dynamics of dark-matter discrete particles be transformed at shorter distance scales to the dynamics of an appropriate classical field?"  In fact there is a significant literature suggesting the possibility of a Bose-Einstein condensate as the dark matter on galactic scales, going back to papers of 1994 by  Sin \cite{sin} and Ji and Sin \cite{jisin} and continuing up to the present.   For a recent review see the work of Lee \cite{lee2}, who was one of the early workers on the subject and continues to explore it today.

If there is one thing on which advocates of (particle) dark matter and advocates of (field-induced) modified gravity seem to agree, it is that if one of these viewpoints is right the other must be wrong.   Caution may be in order about this assertion.    There is a striking precedent:   From the 17th century on, Newton's prestige made the particle hypothesis for light dominant, but early in the 19th century examples of diffraction phenomena overthrew this picture, replacing it with the wave hypothesis.  A hundred years later quantum mechanics showed that both descriptions are needed, each valid in answering appropriate questions.

We observe that there are other proposed modifications of gravity besides MOND.    Mannheim \cite{mannheim} has written an accessible survey of the  subject, beginning with reports in the 1930s of anomalies that could be taken as evidence for dark matter, and including a number of later observational and theoretical works.\footnote{
We have concentrated on MOND because it is the most discussed alternative and, in its original form, amounts to a manifestly successful and economical phenomenology for galactic scales and below.
}

Especially noteworthy is the  work by Mannheim and colleagues  \cite{mannheim} on ``Weyl" or ``conformal" gravity, which uses the symmetry of Weyl's original gauge (i.e., length rather than phase) invariance, alluded to around Eq. (\ref{Wconformal}) \cite{Mannheim:1992tr}.    This is a genuine alternative gravity theory and makes interesting predictions on both  galactic  and longer scales.

In particular, Mannheim has made a prediction based  on conformal gravity that when we learn about the expansion of the universe at still earlier epochs than have been explored up to now we shall find that the expansion already was accelerating.   That statement appears to distinguish this approach from others being considered today, including the most popular ``$\Lambda$CDM" model, i.e., Einstein gravity with dark energy-and cold-dark-matter sources present.

Because conformal gravity involves higher time derivatives, extra boundary conditions are required to make the theory well-defined.   This has led to some debate.  In particular, Flanagan \cite{Flanagan:2006ra} has argued that conformal gravity contradicts the original successful predictions of Einstein gravity for the effects of the Sun.   Mannheim \cite{Mannheim:2007ug} has presented a counter-argument.  We look forward to an eventual consensus on the status of conformal
gravity.\footnote{
A different approach, mentioned briefly by
Mannheim  \cite{mannheim}, has been put forward by Moffat and collaborators, e.g., in \cite{moffat}  and \cite{moffat2}.   
}

Dubovsky et al. \cite{dubovsky2005} produced a development of Rubakov \cite{Rubakov:2004eb}, violating Lorentz invariance and suggesting the possibility of relatively heavy tensor gravitons which might contribute to cold dark matter without Yukawa-like attenuation of static gravitational fields.  This then is a model with both exotic dark matter and modified gravity.  However, an analysis of high-frequency pulsar data by Pshirkov et al. \cite{pshirkov} deduces from the lack of frequency variation that gravitons more or less at rest in the galaxy could not be present in sufficient concentration to make up the dark matter.


\subsubsection{Cluster collisions}

From the viewpoint of testing theories, an unusual object called the ``Bullet Cluster" added important input to the discussion \cite{bullet}. The Bullet Cluster contains two subclusters which appear to have collided some
time ago.   Initially, each subcluster should have contained visible matter in the form of stars, and an order of magnitude more in the form of gas and plasma.   During the collision, the collection of stars should have gone through each other, but the gas clouds from the two subclusters should have experienced a great deal of friction, tending to coalesce and be left behind in the middle.   Dark matter,  being weakly interacting, should have gone straight ahead with little friction or coalescence.  Thus the dark-matter hypothesis leads to an unambiguous prediction about the distribution of matter in the cluster, with the bulk of the matter located near the two star subclusters.

The tool used to analyze this collision was weak gravitational lensing, that is, looking separately with narrow field views in the vicinity of each subcluster of stars, and wide-field views of the whole colliding system, at  week gravitational focusing of light passing by the cluster from more distant sources, and matching this with light images of the system itself in various frequency ranges.  A painstaking analysis showed convincingly that the centers of lensing are located quite near to where the subclusters of galaxies are seen, rather than where the far more massive gas clouds were left behind \cite{bullet}.   The conclusion was powerfully reinforced by an even more demanding analysis studying strong gravitational lensing together with weak lensing \cite{Bradac:2006er}.  This success of a dark-matter prediction is an appealing argument for dark matter on larger distance scales.

Already, certain advocates of modified gravity had acknowledged that they might need some dark matter (possibly in the forms of brown dwarfs and neutrinos of the maximum mass allowed by current limits) to account for the behavior of clusters of galaxies \cite{Milgrom:2008rv}, Thus, the sharp line between the MOND phenomenology on the one hand, and the hypothesis that Einstein gravity is unmodified on the other, became blurred.  This concession by MOND advocates had the consequence that their theory also could predict the kind of lensing pattern observed in the Bullet cluster, so that the result need not distinguish between the two  approaches.

Angus and McGaugh \cite{bulletmond} argue  that a careful analysis of the bullet cluster demonstrates that it is difficult for the CDM model to account for the relative cluster velocity of $\sim 4,700$ km/s determined from the observed shock velocity.  Their argument comes from large scale simulations.  They find it ``difficult (but not unheard of) to achieve $v_{rel}> 4,500$ km/s"  in a CDM scenario, although  they claim that the ``appropriate velocity occurs rather naturally in MOND."  It is important to emphasize that the simulations include only dark matter, and adding ordinary matter would increase gravitational attraction, and hence relative cluster velocity.

While the statements about CDM are based mainly on numerical simulations starting from the earliest stages of expansion of the universe, those for MOND are based to some extent on simulations, but more on qualitative characteristics of the model, i.e., the longer-range force it assumes.   Granting the consensus that MOND works better on the galactic scale while CDM works better on longer scales, this gives incentive for deeper analysis of cluster relative velocities \cite{bulletV}.

Two new works have shed somewhat conflicting light on this matter.
In another colliding cluster system, Abell 520, a weak-lensing analysis indicated that the bulk of the dark matter is located in the core, close to the gas, rather than being associated with the subclusters of galaxies as in the Bullet case \cite{Mahdavi:2007yp}. It appears that the collision velocity is significantly lower here than for the Bullet, and there may even be more than two colliding systems.
If this new finding were to be sustained after completion of an analysis comparable to that achieved for the Bullet cluster, that would be an indication of more complex dynamics, but not necessarily a reason to favor either MOND or CDM, as both give similar predictions for cluster collisions.
On the other hand, it was found \cite{bulletnew} in the system MACSJ0025.4-1222 that the merger looked very similar to the bullet cluster:  two mass peaks located near the optical galaxies and hot emission coming from between them.  Clearly, the question is not yet settled.\footnote{
A recent analysis of motion of small ``satellite" galaxies around a regular galaxy suggests good agreement with the CDM hypothesis.  This does
not necessarily rule out MOND for the region just outside a galaxy
\cite{Angus:2007tz}, but may extend down to even smaller scales than before the successful reach of CDM \cite{Klypin:2007gc}.
}$^,$\footnote{
A well-balanced overview of the competing approaches (even though from a dark matter advocate) can be found in Ref. \cite{dmM}.
}


\subsubsection{Status of cold dark matter (CDM) versus modified Newtonian dynamics (MOND)}

The story for graviton mass is different  from that for photon mass in two principal aspects.
Arguments that the mass must be zero (or at least no bigger than the inverse of the Hubble radius) are much stronger than even what we have just related for photon mass.   On the other hand, there {\it are} observations that can not be reconciled with unadorned GR unless there are new forms of matter, described as dark matter  and dark energy.  Thus, if we generalize the notion of graviton mass to that of long-distance, low-frequency modifications of gravity, then there may indeed be modifications, even though (as we have seen) there are powerful
arguments in favor of extra sources rather than changed 
gravity.\footnote{Actually,
for dark energy the choice between a new source and modification of gravity is not necessarily well-defined.   Einstein's cosmological constant was in his own eyes a modification of gravity.   Even so, inflation models involve a scalar field whose vacuum energy drives exponentially rapid expansion, and such a dynamical cosmological term is placed most naturally on the `matter' side of the Einstein equations.
}

The  dark-matter view faces challenges, of which the biggest is
identifying this matter, which is four or five times bigger than ordinary matter in its contribution to the mass of the universe.  Finding dark matter in the laboratory, perhaps in the form of interactions by dark-matter particles from space incident on sensitive laboratory apparatus, might well settle the issue.
However, it is conceivable that, even if there are dark-matter particles, they are too weakly interacting to be detected by a feasible apparatus.  If so, the case for dark matter would require convincing simulations that reproduce the MOND phenomenology for ordinary-size galaxies.

The main arguments supporting the majority view at the moment are two: \\
\indent 1)   Simulations on the largest scales,  based on both a cosmological term in the Einstein equations (``dark energy") and cold dark matter,  give excellent agreement with a whole set of phenomena.   These include the current baryon-to-photon ratio as well as evidence for  a spatially flat universe with accelerating expansion.  That evidence comes from  three complementary classes of information, data on
i) distant supernovae, ii) the structure of the cosmic microwave background, and  iii) large-scale distributions of galaxies.  \\
\indent 2)  CDM gives a simple and successful foundation for the structure of galaxy clusters, of which the  Bullet Cluster is only the most striking example
As we have seen, ``fixes" already introduced by MOND advocates to explain the gross features of clusters appear to make MOND also compatible with the Bullet results.   Of course the new  report about Abell 520 leaves all of this in possible disarray.
The structure of ordinary galaxies remains the chief open issue for CDM.

Both MOND and CDM in their simplest forms are one-parameter models.   In the case of CDM, that parameter is the mean density of mass constituted by dark-matter particles.   There are several ``hidden" additional parameters.   The mass of such a particle may be exponentially small or exponentially large on the scale of electron volts.  This has little or no effect on the simulations that must reproduce the observed phenomena of the largest scales, such as density fluctuations.  The temperature must be quite low on the scale of the mass, but otherwise its value is immaterial.   The interactions among such particles must be weak, but precise strength also doesn't matter.

For MOND, the one parameter is the critical acceleration $a_0$.  Everything else in galactic-scale phenomenology follows from this.   However, for the extensions embedding MOND into Einstein gravity, more parameters and even functions (new dynamical fields) arise.
Depending on choices for these, the simplest form of MOND might have a cutoff for sufficiently low scales, and the simple rule of a $1/r$ rather than a $1/r^2$ force law might fail at sufficiently large scales.  Thus question marks certainly are present for both approaches as they currently are formulated.

If CDM be right, it seems to us that the simple phenomenology captured by MOND should be open to discovery in more direct ways than by a long evolution of simulations.   The structure looks like an ``attractor," amenable to study by searching for equilibria, stable or at least metastable.  However, the dynamics in this regime may be sensitive to some of the hidden parameters which do not seem to matter
at larger scales or earlier stages in the evolution.   There still may be new physical insights needed to understand in terms of CDM what happens at galactic scales.

MOND starts with a disadvantage, because even though it now has a fully relativistic formulation, this involves more new features than for CDM (which simply introduces another particle joining what already is a substantial collection of particles found during the 20th century).  Nevertheless, theoretical speculations about fields not so different from those found in the generalized MOND have been explored for a long time, making this an intellectually respectable domain for further investigation.   

Our view is that the most satisfying resolution would be one in which basic elements of the two approaches, CDM and MOND can be combined into a single consistent scheme.   At some level, this must be true if physics is to maintain its unity, because both approaches accurately and simply describe broad (and complementary) ranges of phenomena.   Finding an elegant way to unify them (rather than throwing out the ineluctable facts either one describes well) is a major challenge for the field.


\subsection{The challenge for cold dark matter}

Fritz Zwicky \cite{zwicky}, whose report of observational evidence for unexpected accelerations\footnote{He made his observations on the Coma cluster of galaxies.} was the first one not subjected to later serious dispute, may also have been the first to suggest that these accelerations might imply the existence of new phenomena.  In an ambitious book on doing science  \cite{zwicky2} that came out five years before the first edition of Kuhn's \cite{kuhn}, he wrote this about his old observations:   ``{\it A priori} many more hypotheses can be visualized.  Most of these hypotheses are, however of the wild type which we need not consider until all more conventional ideas have proved hopelessly inadequate.  Once this should happen our imagination will be free to experiment with new formulations of the laws of space and gravitation, with the possible variability of the fundamental constants and so on."

The history of the subject actually followed his dictum.   Early thinking about  dark matter was based on the only known  particles that might fit the description -- neutrinos.   These would tend to imply hot dark matter, which soon was ruled out.   Thus the notion of a new weakly-interacting particle became the most conservative candidate hypothesis (or in Zwicky's formulation, the least un-coventional approach) that might explain observations.   

Recent developments have cast into sharp relief both the achievements and the lacunae of the CDM hypothesis as implemented up to now.   We mentioned earlier in this paper that following simulations all the way down to the galactic scale 
with sufficient precision to have some plausability
had been a formidable (and unmet) challenge.   However, now the Via Lactia (Kuhlen et al. \cite{Kuhlen:2008qj}) and Aquarius (Springel et al. \cite{springel}) projects have reported success in achieving this goal.   They find a distribution of dark matter density which has structure at all scales, and far from the galactic center declines as $r^{-3}$, while at small radius it shows a more slowly-varying $r^{-1.2}$. 

There are two big problems with this result, problems that may be related.   1.  We are unaware of any observational evidence for the ``cuspiness" found in the simulation.  2.  Because the simulation does not include baryonic matter, it is inherently unable to account for the Tully-Fisher law, and more fully for the successful MOND phenomenology, which would correspond in terms of dark matter to a density falling as $r^{-2}$, and correlated in a very specific way to the total luminous mass of the galaxy.   Nature seems to be tantalizing us, because weak lensing observations at large radius are compatible with the simulation density distribution $r^{-3}$, but also with the density $r^{-2}$ that would agree with MOND.\footnote{See, for example. Gavazzi et al. \cite{Gavazzi:2007vw}, which in its text is especially favorable to $r^{-2}$, but from its figures appears to tend towards $r^{-3}$ at the larger radii.}

A lack of cuspiness could be due to the influence of the baryonic matter on the dark matter.  The Tully-Fisher law could arise from influence of the dark matter on the amount of baryonic matter, but equally well might involve a mutual interaction between the two types of matter.   In any case, until and unless both types can be taken into account at the same time, the CDM hypothesis will not have confronted, much less passed, a crucial test.   The influence of dark matter on the structure of a galaxy was discussed early on, in particular by Ostriker and Peebles \cite{ospe}. They deduced from a numerical simulation  that an extra attraction beyond the one coming from conventional gravity produced by visible matter alone would speed up motions, and thus tend to stabilize a flat galaxy against the formation of a barred structure.  As only a minority of flat galaxies have a barred structure, this is a pleasing consequence of CDM (perhaps also of MOND).  Given that the amounts of dark and of luminous matter inside the radius of a luminous galaxy should be comparable, it is quite reasonable that the luminous matter would have a significant influence on the distribution of dark matter.   

Pending a satisfactory application of CDM at galactic scales, MOND advocates face their own challenge, because any effort to extend MOND beyond its initial domain of success carries the risk of significant arbitrariness.   Nevertheless, by applying some versions of the critical-acceleration idea at even smaller length scales, such as inside the solar system, one might hope to produce new successful predictions of MOND which would be very hard for CDM to match.   A recent example of this approach may be found in \cite{Milgrom:2009mg}.


\subsection{Out to the far reaches of the universe}

Before even the VdV-Z discontinuity or the Vainshtein nonlinear-gravity effect attenuating this discontinuity comes the prime fact stressed repeatedly in this review : any graviton-mass effect begins with the {\it weakening} of gravitational attraction at long distances.   Galactic rotation curves and motions of objects in clusters of galaxies apparently exhibit exactly the opposite  effect:  a {\it strengthening} of gravity at increasing distances from the center compared to Newtonian expectations based on the visible matter in a system.

Thus, the one thing ``dark matter" could {\it not} be is a graviton-mass effect, where we include in this category generalizations such as the DGP model.   This fact may increase the attractiveness of supposing that some combination of discrete-particle dark matter and continuous-field dark matter might account for these phenomena; in other words, there may be new sources rather than modifications of Newton-Einstein gravity.

Even when the DGP type of approximately zero-mass solution works, it does so only out to a radius
substantially smaller than the effective graviton Compton wavelength
${\lc}_g$.\footnote{Actually,
even for the DGP model, there is effectively a continuum of masses contributing, which  means that the graviton could at best be viewed
as an unstable resonance, certainly not a fixed-mass particle.
}
Because that smaller radius is related to ${\lc}_g$, Gruzinov \cite{gruz}  argued that knowledge about the solar gravitational field implies
\be
{\lc}_g \gtrsim 10^8 ~\mathrm{pc}.
\ee

There is a potentially significant concern here:   The solution is for a source embedded in flat space-time.  However, we live in a world that, although it appears spatially flat, is expanding and even accelerating with the passage of time.  Therefore, it exhibits negative four-dimensional curvature.

The smallest (as well as largest) Compton wavelength it would make sense to consider, if one accepts the objections by Boulware and Deser to Vainshtein's argument, would be about the size of the visible universe, of order
\bea
{\lc}_g  &\gtrsim& 3 \times 10^{26}~\mathrm{m} \equiv 10^{10}~\mathrm{pc} ~~~~~~ \mathrm{or}  \nonumber \\
\mu_g & \lesssim & 6 \times 10^{-32} ~\mathrm{eV}
\equiv 10^{-67} ~\mathrm{kg},
\eea
(meaning exponential growth 
and/or or development of an instability 
would not be substantial).  Even after all the considerations stemming from DGP one is not far from that value.

The conclusion is that nothing except quasi-static fields could be sensitive to a graviton mass, and quite possibly even such fields would not be able to signal such a mass.   Indeed, the consensus even among advocates of the DGP model\footnote{
Gabadadze and Gruzinov  \cite{gruz2} give a nice overview of the reasons to describe a mass-like effect using a higher-dimensional theory.
}
is that the Compton wavelength may be less than infinity but, as remarked by Nicolis and Rattazzi \cite{nicolis}, not appreciably less than the radius of the visible universe.  The reason is quite simple:   A significantly smaller Compton wavelength inevitably would modify drastically phenomena seen on the largest visible scales.   As we discuss later, the DGP model is a possible way of accounting for the accelerating expansion of the universe, but {\it only} with the largest possible Compton wavelength.

However, the accelerating expansion of the universe, indicated by numerous observations in the last decade, {\it is} what one might expect from a weakening of gravity at large distances.   Dvali, Gruzinov and Zaldarriaga \cite{Dvali:2002vf} have considered the possibility  that the length scale corresponding to a DGP `graviton-mass' effect may be comparable with the size of the visible universe, and so could explain quantitatively the observed acceleration.   They then note that there should be small effects at distances within the solar system.\footnote{
The effects are small because of Vainshtein's nonlinear suppression of graviton-mass effects at distances very short compared to the graviton Compton wavelength.
}

In particular, the precession of the perigee\footnote{
The authors refer to `perihelion', but in context it seems clear that `perigee' is intended.
}
of the Moon's orbit\footnote{
In a way, this brings us back to the beginning. In the Principia,  one thing Newton could not calculate satisfactorily was the precession of the Moon's line of apsides.  As described in Ref. \cite{vulcan}, this necessitated two further advances.  The first, was the development of equations of motion for the 3-body problem.   The second was the inclusion of the effects of the Sun's motion with respect to the Earth-Moon system, mainly caused by Jupiter.   (Previously the Sun had been treated as lying at a constant set point.) This was  done in 1749 by the Frenchman Alexis-Claude Clairaut.
}$^,$\footnote{
In 1758 Clairaut applied his perturbation theory to the timing of the return of Halley's comet, with great success \cite{vulcan}.  Thereupon Clairaut became a scion of the Paris salons.  (The French treated scientists well then, as Benjamen Franklin would discover.)   ``Engaged with suppers, late nights, and attractive women, desiring to combine pleasures with his ordinary work, he was deprived of his rest, his health, and finally at the age of only 52, of his life \cite{vulcan}."
}
should have a contribution about an order of magnitude smaller than the sensitivity of current measurements using laser lunar ranging.   It is possible that observational sensitivity could increase sufficiently to detect such a precession.  (The use of lunar laser ranging is discussed in \cite{jdaphgrv}.)

If so, that would tend to confirm a graviton-mass explanation for accelerating expansion, clearly an example of modified gravity.  If such a shift in the perigee precession were ruled out, then that might be an indication instead favoring a modified source, i.e., some form of dark energy.  However, we should hasten to note that there appear to have been no studies of possible dark-energy effects on lunar precession.   There might well be such effects.

Vainshtein's argument for consistency of a non-zero graviton mass  with local observations in non-linear gravity is supported, but the reach of observation to the edge of the visible universe is inconsistent with a graviton Compton wavelength significantly smaller than the radius of that universe, just as asserted by BD in their defense of the VdV-Z argument.   Some of their specific reasoning was refuted by the developments of the DGP model, but as just discussed the BD conclusion nevertheless agrees with current astronomical observations.


\section{Discussion and Outlook}


\subsection{Conclusions}

The subject of possible photon or graviton rest mass is appealing because there are so many levels of beautiful argument for the masses to vanish  (and of course a counter-argument for each argument).  Both of these  examples (of the only long-range fields we know) reveal important strands of physics relevant to many different areas.  Taken as a whole, their study illuminates the history, logic, and remarkably complex yet coherent structure of physics.

In Table \ref{masslist} we give a list of what we find to be the most significant and/or interesting mass limits so far proposed.



\begin{table*}[t!]
\begin{center}
\caption{A list of the most significant mass limits of various types for the photon and graviton.
\label{masslist}} \vskip 20pt
\begin{tabular}{rl|l|l|l|l} \hline\hline
& Description of method ~~~~~~~~~~~~~~ & $\lc \gtrsim$ (m) ~~~~~~ &
  $\mu \lesssim$  (eV) ~~~~~~
  & $\mu \lesssim$ (kg) ~~~~ & Comments  ~~~~~~~~~~~~~~~~~~~~~~    \\   \hline
&                 &             &         &&\\
1 & {\sf Secure photon mass limits:}  && &&\\
&  Dispersion in the ionosphere \cite{Kroll:1971wi} & $8\times 10^5$
 & $3 \times 10^{-13}$ & $ 4\times 10^{-49}$  & \\
& Coulomb's law  \cite{wfh} & $2 \times 10^7$
  &  $10^{-14}$  &  $2 \times 10^{-50}$  &   \\
& Jupiter's magnetic field \cite{DGN}   & $5 \times 10^8$
& $4 \times 10^{-16}$    & $7 \times 10^{-52}$ &  \\
& Solar wind magnetic field \cite{ryutov2}
&  $2 \times 10^{11}$~~~(1.3 AU)
& $10^{-18}$    & $2 \times 10^{-54}$  &
        \\ [10pt]
2 & {\sf Speculative photon mass limits:} &&&&  \\
& Extended Lakes method \cite{lakes,luotorsion,gntorsion} & $3\times10^9$  & $7\times 10^{-17}$  & $10^{-52}$\ &$\lc\sim$ 4 $R_\odot$ to 20 AU,     depending\\
&  &  $~~~~\Leftrightarrow 3\times 10^{12}$
& $~~~~\Leftrightarrow 7 \times 10^{-20}$
& $~~~~\Leftrightarrow 10^{-55}$
& ~~~ on $\mathbf{B}$ speculations \\
& Higgs mass for photon \cite{adel}   &  No limit feasible  ~~
& & &
Strong constraints on 3D Higgs \\
&&&&& ~~~parameter space\\
& Cosmic mag. fields \cite{yamaguchi,chib}, \cite{adel}
& $3\times 10^{19}$  ~~ (${10^3}$ pc)
& $6\times 10^{-27}$ & $ 10^{-62}$
& Needs const. $\mathbf{B}$ in galaxy regions
             \\[10pt]
3 & {\sf Graviton mass limits:}          &&          && \\
& Grav. wave dispersion \cite{finn} & $3 \times 10^{12}$
& $8 \times 10^{-20}$ & $10^{-55}$  &
Question mark for scalar graviton \\
& Pulsar timing \cite{baskaran}
& $2 \times 10^{16}$
& $9 \times 10^{-24}$   & $2 \times 10^{-59}$ &  Fluctuations due to graviton \\
&&&&& ~~~phase velocity \\
& Gravity over cluster sizes  \cite{GNgraviton}
& $2\times 10^{22}$
& $10^{-29}$ & $2 \times 10^{-65}$ & \\
& Near field constraints  \cite{gruz}
& $3 \times 10^{24} ~~(10^8$ pc)
& $6 \times 10^{-32}$ & $10^{-67}$ &
For DGP model \\
&Far field constraints  \cite{Dvali:2002vf}
& $3 \times 10^{26} ~~(10^{10}$ pc)
& $6 \times 10^{-34}$ & $10^{-69}$
&For DGP model
\\[10pt]

\hline\hline
\end{tabular}
\end{center}
\end{table*}


The first reason for vanishing mass is, of course,  something Newton adopted instinctively for gravity and Gauss justified in a beautiful way for electrostatics:  the inverse square law of force.
For gravity this was confirmed with tremendous precision  by the match with Kepler's laws.  For electrostatics Gauss's statement that the number of lines of force coming out of a charge is, at any distance, a direct measure  of that  charge, gave a powerful geometric interpretation to the force law.  (This argument applies equally to mass in gravity.)
Later there came the notions of gauge invariance, discovered in the mathematical structure of classical electrodynamics, and general-coordinate invariance, invented by Einstein to constrain the possible structure of his emerging theory of gravity.

There is also a ``backwards" connection:  As 
Ogievetsky and Polubarinov 
\cite{ogiev1,ogiev2} 
and Weinberg 
\cite{Weinberg:1964ew} 
showed, zero mass for photon or graviton implies local conservation of electric charge in electrodynamics or energy and momentum in gravity.  Like the sizes of the masses, violations of these conservation laws are strongly constrained by experiment.  Thus, for electrodynamics as well as gravity, two effects known to be small are logically related in the limit where they both are zero.

Simplicity also favors these zeros.   They represent the minimal structure
consistent with all symmetry requirements.  Anything different requires more parameters if not more fields.

Despite all these arguments, there is another side to the story.   As Stueckelberg showed, gauge invariance can be satisfied at least formally even in the presence of a mass (and Siegel showed that the same holds for general-coordinate invariance).   Perhaps even more powerful is the example of the W$^\pm$ and Z$^0$ mesons, which  clearly are gauge particles, and yet have mass.

At least for the photon case, there is an objection to this argument.   In the context of a Higgs mechanism one must introduce an electrically charged scalar field, where constraints from observation imply that the value of the charge is an extraordinarily tiny fraction of the charge of an electron.   Such a charge, of course, would violate the pattern of all known charges, and also would contradict an appealing (though unproved) idea, that of grand unified theory.  It would be a bizarre modification of electroweak theory, where the compact group SU(2)$_{\rm left}$ automatically leads to quantized left-handed charges.

In other words, `mini-charged' Higgs particles, if they existed, would perforce be coupled only to the U(1)$_{\rm right}$.  Because weak interactions are short-ranged, and the W and Z bosons are so massive, the effects of the weak charge of the new Higgs particle would be even more insignificant than those of the electric charge.  The effect of the new Higgs coupling on the masses of the W and Z also would be unobservably small.\footnote{
If instead of a Higgs particle, the associated Higgs field were a composite of other fields, then these too would have extraordinarily small, unquantized U(1)$_{\rm right}$ charges, or at least super-small offsets of the charges for different elements of the composite.
}

As far as observation relevant to photon mass goes, the only debate is about how stringent a limit currently can be placed on that mass.   To date no evidence at all has appeared for a nonzero value.  Even with the generalization to a Higgs-mechanism framework, the ``obvious" experiment of seeking to detect dispersion of velocity   with frequency is guaranteed to give no useful information, because limits from static magnetic fields are so low that nothing could be detected by dispersion measurements (at least in regions identified so far where we could measure the velocity).   Even the Schumann resonances do not give as strong a constraint as the magnetostatic limit. Thus, the only even potentially observable effect of photon mass would be found in the photon Compton wavelength, and that already is known to be at least comparable in dimension to the Earth-Sun distance.   Therefore, any future improvements in the limit probably will come from astronomical observation rather than laboratory (even satellite-laboratory) experiment.


\subsection{Prospects}

Let us accept  that the primary tool for limiting or detecting a rest mass of the photon or graviton is to exploit the ``Yukawa" effect:  modification at long distances of essentially static electromagnetic or gravitational fields.  Then  possibilities for extending the range in the case of electromagnetism look very good.  With evolving instruments and techniques, we are in an era of rapid expansion in the depth of exploration of the universe. Detailed knowledge of galactic and extragalactic magnetic fields is  accumulating along with knowledge about the structure of the associated plasmas.   There is every reason to suppose that a lower limit on the photon Compton wavelength limit of galactic or even larger dimensions could  be attainable.

If a finite value for $\lc$ were detected, almost certainly it would be so large and the corresponding photon mass so small that even in the Higgs framework the corresponding electric ``mini-charges" would be too small to detect.  Thus the continuity of electrodynamics in the zero-mass limit (unless electric charge is not locally conserved) already assures that the {\it only} mass effect still possible to observe would be long-distance modifications of static magnetic fields.

In other words, for all lab-scale purposes the mass already may be taken as zero.   Still, if a non-zero value were established by new astronomical
observations, this small departure would have enormous conceptual implications, giving incentive for searching examination of the accepted foundations of electrodynamics.

At the same time, 
the great debate about the possibility of a massive graviton in GR seems pretty much complete.  The most conservative lower bound on the graviton Compton wavelength, based on limits to deviations from Einstein-Newton gravity in the solar system, puts it at $\approx$ 1\% of the radius [R] of the visible universe \cite{gruz}.  Even that estimate is for an asymptotically flat space time, whereas the actual universe is expanding and even accelerating in its 
expansion.\footnote{
So, if spatially flat, the universe still has (negative) curvature in the relation between time and space variables.  This curvature introduces a ghost in the DGP theory, making it suspect, and therefore making the limit \cite{gruz} possibly too conservative.
}
More to the point, a $\lc$ significantly smaller than R would produce dramatic (and certainly not seen) 
effects on the large-scale picture of the universe.

On the issue of dark matter versus modified gravity, we have seen that the most conservative way of accounting for phenomena, the existence of one new weakly interacting particle, has not so far been shown to work for galactic scales, where MOND gives a very successful parametrization of the data.  In our opinion the biggest current challenge for the CDM hypothesis is to remedy this lack, perhaps by finding equilibria involving dark matter and ordinary matter.  If these equilibria implied metastable configurations explaining  flat rotation curves and the Tully-Fisher law, one would have a satisfying confirmation of CDM.
As long as that hasn't happened,  it remains a viable possibility that explaining the MOND phenomenology requires new physics other than CDM.  

Classical electrodynamics and general relativity were the first two field theories of physics.   They appeared  in complete form at the hands of Maxwell and of Einstein, and even today there is no proof that they need to be changed.   Still, remembering that physics is an experimental science, we should keep  watch for  surprises.  It could turn out that either or both of these theories actually require adjustment, perhaps along the lines considered in this paper, or perhaps in some completely new direction.


\begin{acknowledgments}

To begin, it is most appropriate to acknowledge T. Alexander Pond. It was his informal lecture at Stony Brook on 12 June 1968 \cite{mmncline}, which was the original inspiration for our collaboration.\footnote
{It also  seems fitting to bashfully acknowledge Edwin Salpeter.  On 20 October 1961 he gave an electrodynamics exam at Cornell which dealt with solving what are the Proca equations.
MMN absolutely blew it.   
On 13 March 1972 MMN returned to Cornell to gave a talk on the topic pf the photon mass.  At the end he stated that he hoped Ed thought he had done better this time.  (Two of the faculty observed that this was the biggest case of overcompensation they had ever seen.)
}

Further thanks are due to John Anderson, Nima Arkani-Hamed, Jacob Bekenstein, Maru\v sa Brada\v c, Shantanu Desai, Bruce Draine, Sergei Dubovsky, Georgi Dvali, Glennys Farrar, \'Eanna Flanagan, Alexander Friedland, Gregory Gabadadze, George Gillies, Donald Groom,  Andrei Gruzinov, Marek Karliner, Philipp Kronberg, Michael Kuhlen, Rachel Mandelbaum, Philip Mannheim,  Stacy McGaugh, Patrick Meade, Mordehai Milgrom, Lyman Page, Ari Pakman, Gilad Perez, Massimo Porrati, Joel Primack, Dmitri Ryutov, Robert Shrock, Warren Siegel, David Spergel, Volker Springel, George Sterman, Scott Tremaine, Virginia Trimble, Arkady Vainshtein, Cumrun Vafa, Peter van Nieuwenhuizen, Eli Waxman,  Chen Ning Yang, and Hongsheng Zhao.  While their insights have saved us from many mistakes, the ones that remain are our responsibility.

The work of ASG was supported in part by the National Science Foundation.  That of MMN was supported by the United States Department of Energy.

\end{acknowledgments}




\end{document}